\documentclass{emulateapj}

\usepackage{float}
\usepackage{amsmath}
\usepackage{epsfig,floatflt}

\usepackage{graphicx}
\usepackage[]{subfigure}
\usepackage{color}
\definecolor{Blue}{rgb}{0.1,0.1,1.0} 
\definecolor{Magenta}{rgb}{1.0,0.1,0.5} 
\definecolor{LRed}{rgb}{0.8,0.0,0.0}

\newcommand{\nc}{\newcommand}

\nc{\be}[1]{\begin{equation}\mbox{$\label{#1}$}}
\nc{\bea}[1]{\begin{eqnarray} \mbox{$\label{#1}$}}
\nc{\Section}[2]{\section{#2}\label{#1}}
\nc{\Bibitem}[1]{\bibitem{#1}}
\nc{\Label}[1]{\label{#1}}

\nc{\eea}{\end{eqnarray}}
\nc{\ee}{\end{equation}}

\nc{\bdm}{\begin{displaymath}}
\nc{\edm}{\end{displaymath}}
\nc{\dpsty}{\displaystyle}
\nc{\bc}{\begin{center}}
\nc{\ec}{\end{center}}
\nc{\ea}{\end{array}}
\nc{\bab}{\begin{abstract}}
\nc{\eab}{\end{abstract}}
\nc{\btab}{\begin{tabular}}
\nc{\etab}{\end{tabular}}
\nc{\bit}{\begin{itemize}}
\nc{\eit}{\end{itemize}}
\nc{\ben}{\begin{enumerate}}
\nc{\een}{\end{enumerate}}
\nc{\bfig}{\begin{figure}}
\nc{\efig}{\end{figure}}

\nc{\arreq}{&\!=\!&}
\nc{\arrmi}{&\!-\!&}
\nc{\arrpl}{&\!+\!&}
\nc{\arrap}{&\!\!\!\approx\!\!\!&}
\nc{\non}{\nonumber}

\def\lsim{\; \raise0.3ex\hbox{$<$\kern-0.75em
      \raise-1.1ex\hbox{$\sim$}}\; }
\def\gsim{\; \raise0.3ex\hbox{$>$\kern-0.75em
      \raise-1.1ex\hbox{$\sim$}}\; }

\nc{\DOT}{\hspace{-0.08in}{\bf .}\hspace{0.1in}}
\nc{\Laada}{\hbox {$\sqcap$ \kern -1em $\sqcup$}}
\nc\loota{{\scriptstyle\sqcap\kern-0.55em\hbox{$\scriptstyle\sqcup$}}}
\nc\Loota{{\sqcap\kern-0.65em\hbox{$\sqcup$}}}
\nc\laada{\Loota}
\nc{\qed}{\hskip 3em \hbox{\BOX} \vskip 2ex}

\nc{\real}{{\rm I \! R}}
\nc{\Z}{{\sf Z \!\!\! Z}}
\nc{\complex}{{\rm C\!\!\! {\sf I}\,\,}}
\def\bigid{\leavevmode\hbox{\small1\kern-3.8pt\normalsize1}}
\def\id{\leavevmode\hbox{\small1\kern-3.3pt\normalsize1}}
\nc{\slask}{\!\!\!/}
\nc{\bis}{{\prime\prime}}
\nc{\pa}{\partial}
\nc{\ra}{\rangle}
\nc{\goto}{\rightarrow}
\nc{\swap}{\leftrightarrow}

\nc{\EE}[1]{ \mbox{$\cdot10^{#1}$} }
\nc{\abs}[1]{\left|#1\right|}
\nc{\at}[2]{\left.#1\right|_{#2}}
\nc{\norm}[1]{\|#1\|}
\nc{\abscut}[2]{\Abs{#1}_{\scriptscriptstyle#2}}
\nc{\vek}[1]{{\rm\bf #1}}
\nc{\integral}[2]{\int\limits_{#1}^{#2}}
\nc{\inv}[1]{\frac{1}{#1}}
\nc{\dd}[2]{{{\partial #1}\over{\partial #2}}}
\nc{\ddd}[2]{{{{\partial}^2 #1}\over{\partial {#2}^2}}}
\nc{\dddd}[3]{{{{\partial}^2 #1}\over
    {\partial #2 \partial #3}}}
\nc{\dder}[2]{{{d #1}\over{d #2}}}
\nc{\ddder}[2]{{{d^2 #1}\over{d {#2}^2}}}
\nc{\dddder}[3]{{d^2 #1}\over
    {d #2 d #3}}
\nc{\dx}[1]{d\,^{#1}x}
\nc{\dy}[1]{d\,^{#1}y}
\nc{\dz}[1]{d\,^{#1}z}
\nc{\dl}[1]{\frac{d\,^{#1}l}{(2\pi)^{#1}}}
\nc{\dk}[1]{\frac{d\,^{#1}k}{(2\pi)^{#1}}}
\nc{\dq}[1]{\frac{d\,^{#1}q}{(2\pi)^{#1}}}

\nc{\bfT}{{\bf T }}

\nc{\cA}{{\cal A}}
\nc{\cB}{{\cal B}}
\nc{\cD}{{\cal D}}
\nc{\cE}{{\cal E}}
\nc{\cG}{{\cal G}}
\nc{\cH}{{\cal H}}
\nc{\cL}{{\cal L}}
\nc{\cO}{{\cal O}}
\nc{\cT}{{\cal T}}
\nc{\cN}{{\cal N}}
\nc{\cR}{{\cal R}}

\nc{\rvac}[1]{|{\cal O}#1\rangle}
\nc{\lvac}[1]{\langle{\cal O}#1|}
\nc{\rvacb}[1]{|{\cal O}_\beta #1\rangle}
\nc{\lvacb}[1]{\langle{\cal O}_\beta #1 |}
\nc{\bb}{\bar{\beta}}
\nc{\bt}{\tilde{\beta}}
\nc{\ctH}{\tilde{\cal H}}
\nc{\chH}{\hat{\cal H}}
\nc{\al}{\alpha}
\nc{\g}{\gamma}
\nc{\Del}{\Delta}
\nc{\e}{\textrm{e}}
\nc{\eps}{\epsilon}
\nc{\lam}{\lambda}
\nc{\Om}{\Omega}
\nc{\ve}{\varepsilon}
\nc{\mn}{{\mu\nu}}
\nc{\vp}{\varphi}

\nc{\rf}[1]{(\ref{#1})}
\nc{\nn}{\nonumber \\*}
\nc{\bfB}{\bf{B}}
\nc{\bfv}{\bf{v}}
\nc{\bfx}{\bf{x}}
\nc{\bfy}{\bf{y}}
\nc{\vx}{\vec{x}}
\nc{\vy}{\vec{y}}
\nc{\oB}{\overline{B}}
\nc{\oI}{\overline{I}}
\nc{\oR}{\overline{R}}
\nc{\rar}{\rightarrow}
\nc{\ti}{\times}
\nc{\slsh}{\hskip-5pt/}
\nc{\sm}{Standard~Model~}
\nc{\MP}{M_{\rm Pl}}
\nc{\mpl}{M_{\rm Pl}}
\nc{\tp}{t_{\rm Pl}}

\nc{\pmin}{p_{\rm min}}
\nc{\pmax}{p_{\rm max}}
\nc{\fo}{f_0}
\nc{\foi}{f_{0,i}\,}
\nc{\fop}{f_0^P}
\nc{\fou}{f_0^U}

\nc{\eff}{{\rm eff}}
\nc{\MT}{M_{\rm T}}
\nc{\ML}{M_{\rm L}}
\nc{\kk}{\vek{k}}
\nc{\pp}{{\rm p}}
\nc{\pt}{\partial_t}
\nc{\half}{{1\over 2}}
\nc{\w}{\omega}
\nc{\uhat}{\hat{U}_\w}

\nc{\etal}{\mbox{\it et al.}}
\nc{\ie}{{\it i.e. }}
\nc{\eg}{{\it e.g. }}
\nc{\trh}{T_{\rm RH}}
\nc{\ad}{{a'\over a}}
\nc{\bd}{{b'\over b}}
\nc{\Rd}{{R'\over R}}
\nc{\diag}{{\textrm{diag}}}
\nc{\mato}[1]{\tilde{#1}}
\nc{\sinn}{\textrm{sinn}}
\nc{\sech}{\textrm{sech}}
\nc{\I}{\textrm{I}}
\nc{\II}{\textrm{II}}
\nc{\III}{\textrm{III}}
\nc{\vev}[1]{\langle #1 \rangle}
\nc{\hyp}{\,\; F_{1{\hskip -16pt}2}{\hskip 11pt}}
\nc{\brhom}{\overline{\rho}_M}
\nc{\brho}{\overline{\rho}}
\nc{\rhob}{\overline{\rho}}
\nc{\Pb}{\overline{P}}
\nc{\bH}{\overline{H}}
\nc{\ep}{{1+4\eps}}

\nc{\deriv}[2]{ 
\frac{\mathrm{d}#1}{\mathrm{d}#2}
}
\nc{\Mnu}{M_\nu}
\nc{\bee}{\begin{equation}}
\nc{\ene}{\end{equation}}
\nc{\hdp}{\sigma_8 (\Omega_{\rm m}/0.3)^{0.37}}
\nc{\avis}{\alpha_{vis}}
\nc{\cvis}{c^2_{vis}}
\nc{\clam}{c^2_{lam}}

\def\smiley{\hbox{\large$\bigcirc$\hspace{-.80em}%
\raise.2ex\hbox{$\cdot\cdot$}\kern-.61em    
\lower.2ex\hbox{\scriptsize$\smile$}}\ }

\def\frowney{\hbox{\large$\bigcirc$\hspace{-.80em}%
\raise.2ex\hbox{$\cdot\cdot$}\kern-.635em
\lower.2ex\hbox{\scriptsize$\frown$}}\ }

\begin{document}

\title{Bayesian analysis of sparse anisotropic universe models and application to the 5-yr WMAP data}
\author{Nicolaas E. Groeneboom\altaffilmark{1} and Hans Kristian Eriksen\altaffilmark{1,2}}

\email{nicolaag@astro.uio.no}
\email{h.k.k.eriksen@astro.uio.no}

\altaffiltext{1}{Institute of Theoretical Astrophysics, University of
  Oslo, P.O.\ Box 1029 Blindern, N-0315 Oslo, Norway}

\altaffiltext{2}{Centre of Mathematics for Applications, University of
  Oslo, P.O.\ Box 1053 Blindern, N-0316 Oslo, Norway}

\date{\today}

\begin{abstract} 
  We extend the previously described CMB Gibbs sampling framework to
  allow for exact Bayesian analysis of anisotropic universe models,
  and apply this method to the 5-year WMAP temperature
  observations. This involves adding support for non-diagonal signal
  covariance matrices, and implementing a general spectral parameter
  MCMC sampler. As a worked example we apply these techniques to the
  model recently introduced by Ackerman et al., describing for
  instance violations of rotational invariance during the inflationary
  epoch. After verifying the code with simulated data, we analyze the
  foreground-reduced 5-year WMAP temperature sky maps. For $\ell \le
  400$ and the W-band data, we find tentative evidence for a preferred
  direction pointing towards $(l,b) = (110^{\circ}, 10^{\circ})$ with
  an anisotropy amplitude of $g_* = 0.15 \pm 0.039$. Similar results
  are obtained from the V-band data [$g_* = 0.10\pm 0.04$; $(l,b) =
    (130^{\circ}, 20^{\circ})$]. Further, the preferred direction is
  stable with respect to multipole range, seen independently in both
  $\ell=[2,100]$ and $[100,400]$, although at lower statistical
  significance. We have not yet been able to establish a fully
  satisfactory explanation for the observations in terms of known
  systematics, such as non-cosmological foregrounds, correlated noise
  or asymmetric beams, but stress that further study of all these
  issues is warranted before a cosmological interpretation can be
  supported.
\end{abstract}

\keywords{cosmic microwave background --- cosmology: observations --- 
methods: numerical}

\maketitle

\section{Introduction}   
\label{secintroduction}

Since the early 1990's, great advances have been made in the field of
data analysis techniques for studying the cosmic microwave background
(CMB). Observations of the CMB anisotropies, for instance those made
by the Wilkinson Microwave Anisotropy Probe (WMAP) experiment
\citep{bennett:2003, hinshaw:2007}, provides the single most powerful
probe in contemporary cosmology.  From these, various theoretical
universe models may be constrained, and today an effective concordance
model based on the inflationary $\Lambda$CDM framework has been
established.

The theory of inflation was initially proposed as a solution to the
horizon and flatness problem \citep{guth:1981}. Additionally, it
established a highly successful theory for the formation of primordial
density perturbations, thus providing the required seeds for the
large-scale structures (LSS), later giving rise to the temperature
anisotropies in the cosmic microwave background radiation that we
observe today \citep{starobinsky:1980, guth:1981,linde:1982, muhkanov:1981,
  starobinsky:1982, linde:1983, linde:1994, smoot:1992, ruhl:2003,
  runyan:2003, scott:2003}.  

A firm prediction of inflation is that the observed universe should be
nearly isotropic on large scales. Yet, recent theoretical studies have
demonstrated that anisotropic inflationary models are indeed
conceivable \citep{armendariz:2006, contaldi:2007, pullen:2007, kanno:2008,
  yokoyama:2008}. Two other examples are those presented by
\citet{ackerman:2007} (ACW) and \citet{erickcek:2008}. The first model
considers violation of rotational invariance in the early universe,
while the second model describes the effects on the observed
perturbation distribution due to a large-scale curvaton field. 

The introduction of anisotropic models poses several problems in terms
of data analysis. The definition of a proper likelihood function may
be non-trivial for a general case, although many models can be
described as multivariate Gaussians with non-diagonal covariance
matrices. All models mentioned above are examples of
this. Yet, even in these relatively simple cases, the numerical
evaluation of the likelihood is computationally unfeasible due to the
sheer size of the relevant covariance matrix.

In the present paper, we extend the previously described CMB Gibbs
sampling framework \citep{jewell:2004, wandelt:2004, eriksen:2004b} to
allow for non-diagonal, but sparse, covariance matrices. As currently
described in the literature, this framework allows for exact Bayesian
analysis of high-resolution CMB data, but only under the assumption of
isotropy, i.e., a diagonal CMB covariance matrix. This method has
already been applied several times to the WMAP data
\citep{odwyer:2004, eriksen:2007a, eriksen:2007b, eriksen:2008b}, and
has been extended to take into account both polarization
\citep{larson:2007} and internal component separation
\citep{eriksen:2008a}.

The question of isotropy has received considerable attention during
recent years, due to unexpected signatures observed in the WMAP sky
maps. These data appear to exhibit several significant and distinct
signatures of violation of statistical isotropy. First, \citet{de
  Oliveira-Costa:2004} found a striking alignment between the two
largest harmonic modes in the temperature anisotropy sky, the
quadrupole and the octopole. Second, \citet{vielva:2004} pointed out
the presence of a very large cold spot in the southern Galactic sky,
apparently incompatible with $\Lambda$CDM-based simulations. Finally,
\citet{eriksen:2004a} found a significantly anisotropic distribution
of power between two hemispheres. The tools developed in the present
paper may be able to constrain specific models relevant for these
observations. In particular, we use these methods to estimate the
anisotropy parameters in the ACW model from the 5-year WMAP
temperature data.

The paper is structured as follows: In \S \ref{sec:acwmodel},
we review the ACW universe model, and briefly introduce the relevant
posterior distribution. Next, we present the method in
\S\ref{sec:method}, before we apply our tools to simulated data in
\S\ref{sec:simulations}. In \S\ref{sec:wmap} we analyze
the five-year WMAP temperature sky maps. Finally, we conclude in
\S\ref{sec:conclusions}.

\section{The anisotropic ACW universe model}
\label{sec:acwmodel}

\begin{figure}
    \includegraphics[width=80mm]{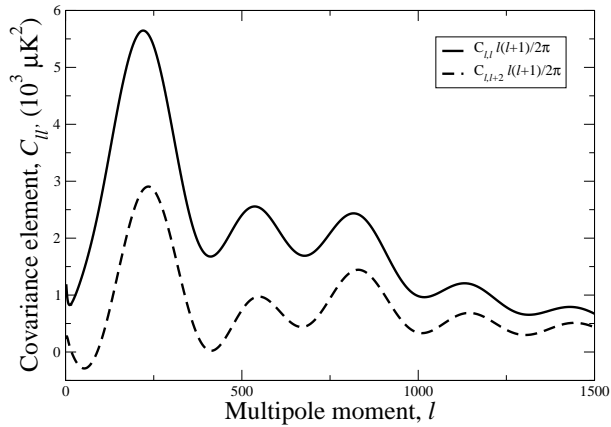}

\caption{Covariance elements, $C_{\ell,\ell}$ and $C_{\ell,\ell+2}$,
  used in the construction of the ACW covariance matrix. These are
  computed by modifying CAMB, a publicly available Boltzmann code.}
\label{fig:covar}
\end{figure}

There has been a surge of interest in anisotropic universe models
since the release of the 1-year WMAP data in 2003, when several hints
of violation of statistical isotropy and/or non-Gaussianity were
reported. One such model was devised by ACW in order to study
violations of rotational invariance during the inflationary epoch.  In
this section, we briefly review this model as it will be used as an
worked example of the general analysis framework. However, we
emphasize that the methods described in this paper are general and
suitable for any universe model that predicts a sparse CMB signal
covariance matrix.

ACW considered breaking of rotational invariance by generalizing the
spectrum of primordial density perturbations $P(k)$ to include a
preferred direction, $\hat{\mathbf{n}}$, as well as wave-number $k$,
\begin{equation}
  P(\mathbf{k}) = P(k) (1 + g(k)(\hat{\mathbf{k}}\cdot\hat{\mathbf{n}})^2).
\end{equation}
Here $\hat{\mathbf{k}}$ is the unit vector along $\mathbf{k}$, and
$g(k)$ is a general function of $k$. Using a combination of
naturalness arguments and detailed analysis of specific models, ACW
then argued that $g(k)$ in most cases can be well approximated by a
simple constant, $g_*$, and presented the full CMB covariance matrix
corresponding to this modified power spectrum,
\begin{equation}
\label{eqcovariance}
  S_{\ell m, \ell' m'} = C_{\ell}\delta_{\ell \ell'} \delta_{m m'} + 
\Delta_{\ell m, \ell' m'}.
\end{equation}
Here $S_{\ell m, \ell' m'} = \langle a_{\ell m} a^*_{\ell' m'}
\rangle$ is the CMB signal covariance matrix, $C_{\ell}$ is the
angular CMB power spectrum given as
\begin{equation}
  C_\ell = \int dk k^2 P(k) \theta_\ell^2(k)
\end{equation}
where $\Theta_\ell(k)$ is the transfer function. The term $\Delta_{\ell m, \ell' m'}$ is then defined
as
\begin{eqnarray}
  \Delta_{\ell m, \ell' m'} &= g_* \xi_{\ell m, \ell' m'} \int_0^\infty
  dk k^2 P(k) \Theta_\ell(k) \Theta_{\ell'}(k) \\
&= g_* \xi_{\ell m, \ell' m'} C_{\ell,\ell'}. 
  \label{eq:delta}
  \nonumber
\end{eqnarray}
In this expression, $\xi_{\ell m, \ell' m'}$ are geometric
coefficients (see ACW for explicit details). The $\xi$ coefficients couple $\ell$ to
$\ell'=\{\ell,\ell\pm2\}$ and $m$ to $m'=\{m, m\pm1, m\pm2\}$. All
other elements are zero.

\begin{figure*}
\includegraphics[width=170mm]{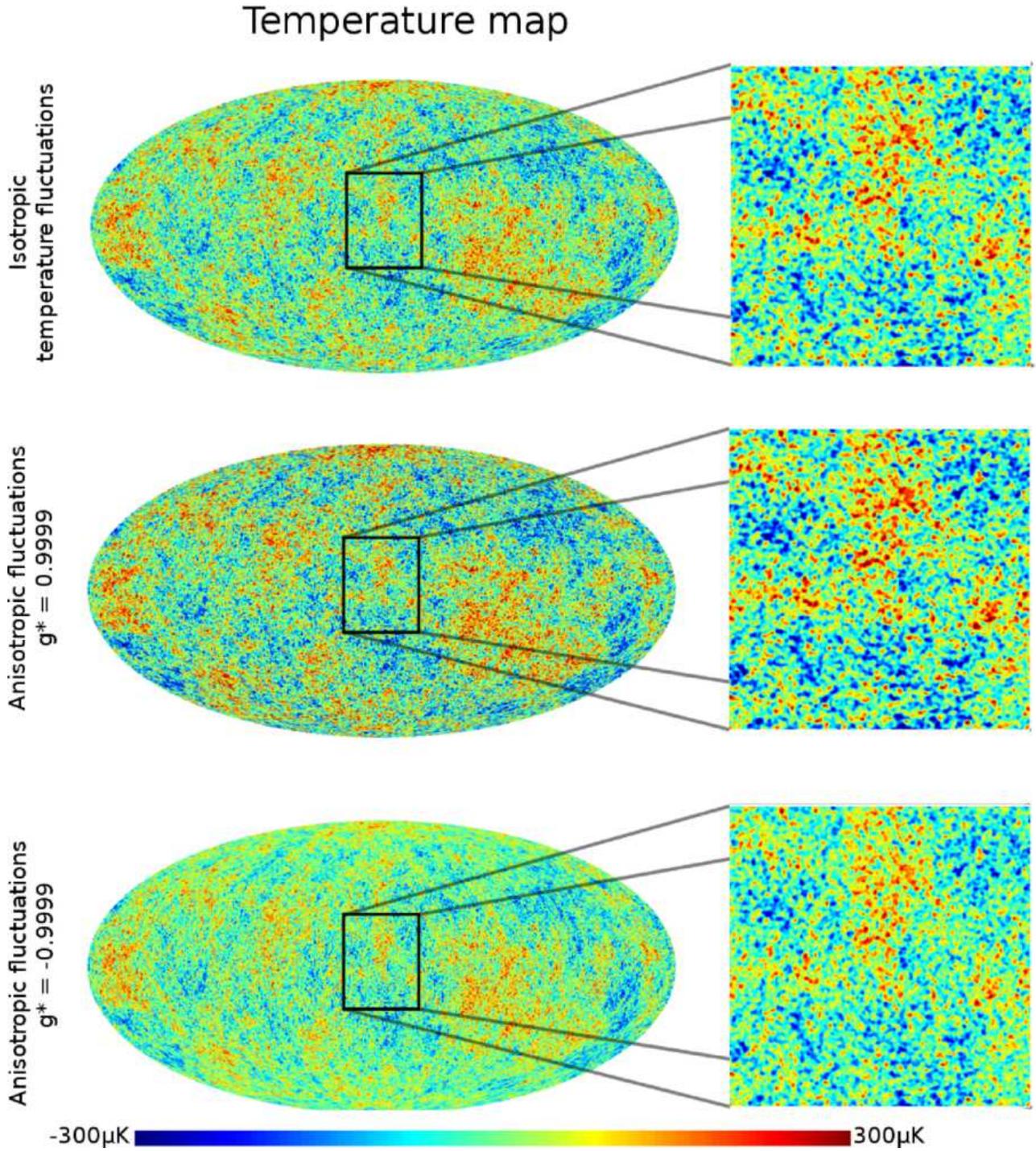}
\caption{Temperature maps showing isotropic fluctuations (top row),
  while the two lower rows depict anisotropic contributions with
  $g_*=0.9999$ (middle row) and $g_*=-0.9999$ (bottom row). The maps
  in the left column are presented in Mollweide projection, while the
  right row is Cartesian.  The anisotropy direction was chosen to be
  $(l,b) = (0^{\circ}, 90^{\circ})$. Note the subtle tendency for
  stripes along the equator for the positive $g_*$, and perpendicular
  to the equator for negative $g_*$. }
\label{fig:temperaturemaps}
\end{figure*}

The coupling to standard cosmological parameters enter only through
$C_{\ell,\ell'}$, which is a straightforward generalization of the
angular CMB power spectrum. In this paper, we assume that the
cosmological parameters are known, and only the anisotropy parameters,
$g_*$ and $\hat{\mathbf{n}}$, are unknown. We therefore compute
$C_{\ell,\ell'}$ once, using a very slightly modified version of CAMB
\citep{lewis:2000} that outputs $C_{\ell,\ell+2}$ in addition to
$C_{\ell,\ell}$, and adopt this matrix as a prior. We adopt the
best-fit $\Lambda$CDM model determined from the 5-year WMAP data
\citep{komatsu:2008}, and the corresponding $C_{\ell,\ell}$ and
$C_{\ell,\ell+2}$ elements are plotted in Figure
\ref{fig:covar}. Joint estimation of cosmological parameters and the
anisotropy parameters will be considered in a future publication.

In Figure \ref{fig:temperaturemaps} we show one realization drawn from
a Gaussian distribution with zero mean and $S_{\ell m, \ell'm}$ as
covariance matrix, with $g_* = 0.9999$ (middle row) and $g_* =
-0.9999$ (bottom row) and a preferred direction of $(l,b) = (0^\circ,
90^\circ)$. The isotropic signal is depicted in the top row. 

The anisotropic contribution alone consists of correlations with the
underlying isotropic signal stretched along the plane normal to the
preferred direction. The sign of $g_*$ determines whether the
anisotropic contribution is to be added or subtracted from the
isotropic signal. If the anisotropic signal is added, then the spots
are stretched along the plane normal to the preferred direction.
However, if the anisotropic signal is subtracted ($g_*<0)$, then the
spots are effectively squeezed along the plane normal to the preferred
direction, corresponding to stretching parallel to the preferred
direction.

\subsection{The  $A_{s}$-- $g_{*}$ degeneracy}
\label{sec:degeneracy}

From equation (\ref{eqcovariance}) and the definition of $\xi$ (see ACW)
it is clear that $\Delta$ contributes also to the diagonal of the
signal covariance matrix, and therefore affects the total angular
power spectrum, not only the correlations among $a_{\ell m}$'s. This
introduces a strong degeneracy between $g_*$ and the amplitude of the
power spectrum of scalar perturbations, $A_s$ or $\sigma_8$. Unless
one attempts to estimate the standard $\Lambda$CDM parameters jointly
with the new anisotropy parameters, one must therefore ensure that a
given choice of $g_*$ does not significantly affect the overall power
spectrum, but only the anisotropic contribution.

The diagonal part of $\Delta$, for which the integral over the
transfer functions equals $C_{\ell}$, is
\begin{equation}
\Delta_{\ell m, \ell m} = g_* C_{\ell} \xi_{\ell m, \ell m}.
\end{equation}
The net extra power due to $\Delta$ is therefore
\begin{equation}
D_{\ell} = \frac{g_*C_{\ell}}{2\ell+1} \sum_{m=-\ell}^{\ell} \xi_{\ell m,
  \ell m}.
\end{equation}
This may be greatly simplified by considering the detailed form of
$\xi_{\ell m, \ell m}$,
\begin{equation}
\xi_{\ell m, \ell m} = -2 n_+ n_-
\frac{-1+\ell+\ell^2+m^2}{(2\ell-1)(2\ell+3)} +
 n_0^2
\frac{2\ell^2+2\ell-2m^2-1}{(2\ell-1)(2\ell+3)},
\end{equation}
where
\begin{equation}
n_+ = -\frac{n_x -in_y}{\sqrt{2}}; n_- = \frac{n_x+in_y}{\sqrt{2}};
n_0 = n_z.
\end{equation}
Averaging this expression over $m$, one finds that 
\begin{equation}
\frac{1}{2\ell+1} \sum_{m=-\ell}^{\ell} \xi_{\ell m,
  \ell m} = \frac{1}{3},
\end{equation}
such that $D_{\ell} = \frac{1}{3} g_* C_{\ell}$. We therefore redefine
the total signal covariance matrix to read
\begin{equation}
S_{\ell m, \ell' m'} = \frac{C_{\ell}\delta_{\ell \ell'} \delta_{m m'} + 
\Delta_{\ell m, \ell' m'}, }{1+g_*/3}.
\label{eq:covar_mat}
\end{equation}
With this definition, $g_*$ is a direct measure of the anisotropic
component of $\mathbf{S}$, and does not directly depend on the power
spectrum $C_{\ell}$.

The effect of $g_*$ on the power spectrum is demonstrated in Figure
\ref{fig:rescaling}, where we plot the power spectra of a simulated
anisotropic map with $g_*= 3$, with and without the above
rescaling. Unless proper rescaling is performed, or some equivalent
parametrization introduced, it is clear that the strongest constraints
on $g_*$ will come from the observed power spectrum, rather than the
correlations among $a_{\ell m}$'s.

\subsection{Posterior analysis and priors}
\label{sec:priors}

The goal is now to estimate $g_*$ and $\hat{\mathbf{n}}$ from observed
CMB maps, by computing the posterior distribution $P(g_*,
\hat{\mathbf{n}}|\mathbf{d})$, $\mathbf{d}$ denoting the data. Because
we assume that both the noise and CMB sky signal are Gaussian (but
anisotropic) random fields, this distribution reads, by Bayes' theorem,
\begin{equation}
  P(g_*, \hat{\mathbf{n}}|\mathbf{d}) \propto \mathcal{L}(g_*, \hat{\mathbf{n}}) P(g_*, \hat{\mathbf{n}}),
\end{equation}
where $\mathcal{L}(g_*, \hat{\mathbf{n}}) = P(\mathbf{d}|g_*, \hat{\mathbf{n}})$ is the likelihood
\begin{equation}
\label{eq:likelihood1}
  \mathcal L(g_*, \hat{\mathbf{n}}) \propto \frac{e^{-\frac{1}{2}\mathbf{d}^T
      \mathbf{C}^{-1}\mathbf{d}}}{\sqrt{|\mathbf{C}|}}
\end{equation}
and $P(g_*, \hat{\mathbf{n}})$ is a prior. 
Equation (\ref{eq:likelihood1}) can be evaluated in $\mathcal O(N_{\textrm{pix}}^2)$ operations, as shown by \cite{oh:1999}.
In this expression,
$\mathbf{C}$ is the signal-plus-noise covariance matrix. In principle,
we could now simply map this distribution over a three-dimensional
grid, and our task would be completed. However, except for the special
case of a data set with uniform noise and full-sky coverage, this is
in practice impossible because $\mathbf{C}$ is a dense matrix, and
inversion and matrix determinant therefore scales as
$\mathcal{O}(N_{\textrm{pix}}^3)$, $N_{\textrm{pix}}$ being the number
of pixels. For current and future data sets, one expects
$N_{\textrm{pix}}\sim 10^6$ or more.

\begin{figure}
\includegraphics[width=85mm]{f3.eps}
\caption{Power spectra of simulated anisotropic sky maps with $g_* =
  3$, with (green) and without (black) rescaling. Red curve shows the
  power spectrum for an isotropic simulation with $g_* = 0$.}
\label{fig:rescaling}
\end{figure}

\begin{figure}
\includegraphics[width=80mm]{f4b.eps}
\includegraphics[width=80mm]{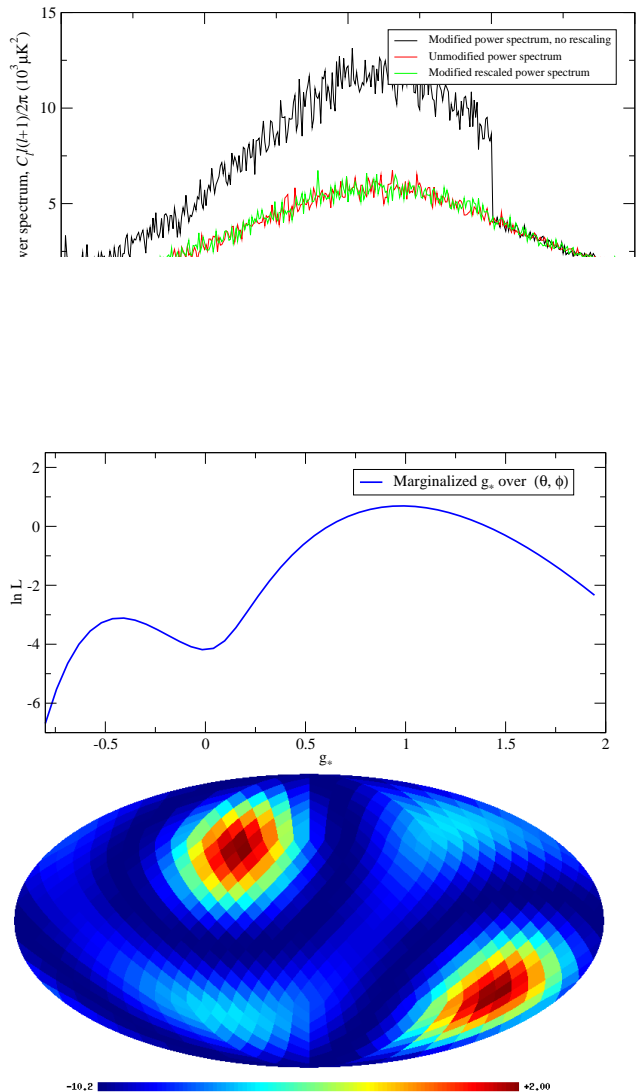}
\caption{Marginal likelihood functions for $\mathcal{L}(g_*)$ (top)
  and $\mathcal{L}(\hat{\mathbf{n}})$ (bottom) for a simulated data set with
  uniform noise and full-sky coverage, shown in logarithmic units. The
  input values of $g_* = 0.8$ and $(l,b) = (57^{\circ}, 33^{\circ})$
  are accurately reproduced. Notice the shallow local maximum at $g_*
  \sim -0.5$ and the secondary peaks in the marginal direction map.}
\label{fig:brute_loglik}
\end{figure}

Fortunately, there is one specific feature of the ACW model that does
make an exact analysis possible: Although the \emph{full-sky} CMB
covariance matrix is non-diagonal, it is not dense. Rather, it has a
well-defined shape in harmonic space ($\ell$ is coupled to
$\ell'=\{\ell,\ell\pm2\}$ and $m$ to $m'=\{m, m\pm1, m\pm2\}$) that
allows for cheap matrix storage and fast Cholesky decomposition. This,
combined with the development of the standard diagonal CMB Gibbs
sampler mentioned in the introduction \citep{jewell:2004,
  wandelt:2004, eriksen:2004b}, allows us to perform a full proper
analysis, as explained in the next section.

Before describing this method, we consider first the special case of
data having uniform noise and full-sky coverage, which is useful to
illustrate the approach, and highlight some particular issues. For
this particular case, the full data covariance matrix, expressed in
spherical harmonic space, has the same sparse filling pattern as the
ACW covariance matrix, and direct evaluation is therefore possible
using sparse matrix techniques \citep[e.g.,][]{tdavis:2005}.

We simulated a single CMB realization from the ACW model, adopting a
high anisotropy amplitude of $g_* = 0.8$ and a preferred direction
(in Galactic longitude and latitude) of $(l,b) = (57^{\circ},
33^{\circ})$, then convolved this realization with a $90'$ FWHM
Gaussian beam, and projected it onto a
HEALPix\footnote{http://healpix.jpl.nasa.gov} grid with resolution
parameter $N_{\textrm{side}} = 128$. Finally, uniform, Gaussian noise
with $10 \mu\textrm{K}$ RMS was added to each pixel. This simulation
was then analyzed by computing the raw likelihood over a
three-dimensional grid, and finally marginalized likelihoods were
produced by numerical integration. The results from this exercise is
shown in Figure \ref{fig:brute_loglik}.

As expected, the likelihood peaks close to the input values. However,
there is also a second local maximum at $g_* \sim -0.5$ with a
direction of $(l,b) \sim (45^{\circ},-50^{\circ})$, $90^{\circ}$ with
respect to the main axis. This maximum becomes visible only for large
negative values of $g_*$. The existence of this maximum becomes
intuitive when considering figure \ref{fig:temperaturemaps}: Flipping
the sign of $g_*$ and rotating the preferred axis by $90^{\circ}$ leads
to stripes in the same direction as the original parameters.

This is not a significant issue for a direct evaluation method, since
the local maximum has a very small amplitude.  (Note that the marginal
likelihoods in Figure \ref{fig:brute_loglik} are shown in logarithmic
units.) However, for MCMC methods it can cause problem in terms of
burn-in: As explained in the next section, our method is based on the
well-known MCMC and Gibbs sampling algorithms, and these essentially
correspond to performing a random walk on the likelihood
surface. Further, each chain is initialized randomly on the sphere.
It is therefore a significant chance that a number of chains may get
trapped in a local maximum, and thereby bias the final posterior. To
avoid this, we impose a uniform prior of $g_* \ge -0.2$ in this paper,
and a uniform prior on the sphere for $\hat{\mathbf{n}}$. If the final
posteriors from the actually WMAP analysis happened to peak close to
$g_* = -0.2$ we would have to re-consider this choice more carefully,
but as we shall see, this is not the case.

\section{Method}
\label{sec:method}

We now discuss the method for mapping out the desired posterior. This
method is a very slight generalization of the previously described CMB
Gibbs sampler developed by \citet{jewell:2004}, \citet{wandelt:2004}
and \citet{eriksen:2004b}, which was originally intended for power
spectrum estimation. The underlying Gibbs sampler implementation used
for this work is the code called ``Commander'', described in detail by
\citet{eriksen:2004b, eriksen:2008a}.

\subsection{Review of the CMB Gibbs sampler}

We first review the CMB Gibbs sampler as previously described in
literature. In any Bayesian analysis, a main goal is the posterior
distribution $P(\theta | \mathbf{d})$, where $\theta$ is a set of
parameters connected to some model and $\mathbf{d}$ are the observed
data.  For high-dimensional spaces, brute-force evaluations of the
posterior are computationally unfeasible, and one usually resorts to
Monte Carlo Markov chain (MCMC) methods. 

\subsubsection{Notation and data model}

We begin by defining a parametric model for the CMB
observations. Given our current understanding of the CMB sky, the
observed data may be accurately modelled as a sum of a CMB anisotropy
term and a noise term,
\begin{equation}
\mathbf{d}  = \mathbf{A}\mathbf{s} + \mathbf{n}.
\end{equation}
Here $\mathbf{d}$ represents the observed data, $\mathbf{A}$ denotes
convolution by an instrumental beam, $\textbf s(\theta,\phi) =
\sum_{\ell,m} a_{\ell m} Y_{\ell m}(\theta,\phi)$ is the CMB sky
signal represented in either harmonic or real space, and $\textbf n$
is instrumental noise.

Further, it is a good approximation to assume both the CMB and noise
to be zero mean Gaussian distributed variates, with covariance
matrices $\mathbf{S}$ and $\mathbf{N}$, respectively. In harmonic
space, the signal covariance matrix is defined by $\textbf S_{\ell
  m,\ell'm'} = \left< a_{\ell m} a_{\ell 'm'}^* \right>$, which may or
may not be diagonal. The connection to cosmological parameters
$\theta$ is made through this covariance matrix.  Finally, for
experiments such as WMAP, the noise is often assumed uncorrelated
between pixels, $\textbf N_{ij} = \sigma_{i}^2 \delta_{ij}$, for
pixels $i$ and $j$, and noise RMS equals to $\sigma_{i}$.

Our goal is now to compute the full joint posterior $P(\theta |
\mathbf{d})$, which, as already mentioned, is given by $P(\theta
|\mathbf{d}) \propto P(\mathbf{d}| \theta ) P( \theta) = \mathcal{L}(
\theta) P(\theta),$ where $\mathcal{L}( \theta )$ is the likelihood,
and $P(\theta)$ is a prior. For a Gaussian data model, the likelihood
is
\begin{equation}
  \mathcal L(\theta) \propto \frac{e^{-\frac{1}{2}\mathbf{d}^T
      \mathbf{C}^{-1}(\theta)\mathbf{d}}}{\sqrt{|\mathbf{C(\theta)}|}}.
\label{eq:likelihood}
\end{equation}

\subsubsection{Posterior mapping by Gibbs sampling}

When working with real-world CMB data, there are a number of issues
that complicate the analysis. Two important examples are anisotropic
noise and Galactic foregrounds. First, because of the scanning motion
of a CMB satellite, the pixels in a given data set are observed by
unequal amounts of time. This implies that the effective noise is a
function of position on the sky. Second, large regions of the sky are
obscured by Galactic foregrounds (e.g., synchrotron, free-free and
dust emission), and these regions must be rejected from the analysis
by masking.

Because of these issues, the total data covariance matrix
$\mathbf{S}+\mathbf{N}$ is dense in both pixel and harmonic space. As
a result, it is computationally difficult to evaluate the likelihood
in Equation (\ref{eq:likelihood}), since the computational cost of
matrix inversion and determinant evaluation scale as
$\mathcal{O}(N_{\textrm{pix}}^3)$.  Fortunately, this problem has
already been solved for the CMB context, through the development of
the CMB Gibbs sampler.

The idea behind the CMB Gibbs sampler is to estimate the CMB sky,
$\mathbf{s}$, together with the covariance parameters, by computing
$P(\theta,\mathbf{s}|\mathbf{d})$, and then subsequently marginalize
over $\mathbf{s}$. Specifically, the algorithm is the following: First
choose any initial guess, $(\theta, \mathbf{s})^0$. Then alternately
sample from each of the \emph{conditional} distributions,
\begin{eqnarray}
\theta^{i+1} \leftarrow& P(\theta| \mathbf{s}^i,
  \mathbf{d}) \\
\mathbf{s}^{i+1} \leftarrow& P(\mathbf{s} | \theta^{i+1}, \mathbf{d}).
\end{eqnarray}
The theory of Gibbs sampling then guarantees that the joint samples
$(\theta, \mathbf{s})^i$ will, after some burn-in period, be drawn
from the desired joint distribution. The remaining step is then simply
to formulate sampling algorithms for each of the two conditionals,
$P(\theta| \mathbf{s}, \mathbf{d})$ and $P(\mathbf{s} | \theta,
\mathbf{d})$.

We first consider $P(\mathbf{s} | \theta, \mathbf{d})$. This may,
under the assumption of Gaussianity, be written as
\begin{eqnarray}
P(\mathbf{s} | \theta, \mathbf{d}) &= P(\mathbf{d}|\mathbf{s}, \theta)
P(\mathbf{s}|\theta) \\
&\propto e^{-\frac{1}{2} (\mathbf{d}-\mathbf{s})^T \mathbf{N}^{-1}
    (\mathbf{d}-\mathbf{s})} e^{-\frac{1}{2}\mathbf{s}\mathbf{S}^{-1}
    \mathbf{s}} \\
  & = e^{-\frac{1}{2}(\mathbf{s}-\hat{\mathbf{s}})^T (\mathbf{S}^{-1}
    + \mathbf{N}^{-1}) (\mathbf{s}-\hat{\mathbf{s}})},
\end{eqnarray}
where we have defined the Wiener filtered map, $\hat{\mathbf{s}} =
(\mathbf{S}^{-1} +
\mathbf{N}^{-1})^{-1}\mathbf{N}^{-1}\mathbf{d}$. Thus, $P(\mathbf{s} |
\theta, \mathbf{d})$ is a Gaussian distribution with mean
$\hat{\mathbf{s}}$ and covariance $(\mathbf{S}^{-1} +
\mathbf{N}^{-1})^{-1}$.

Sampling from this distribution is straightforward, but
implementationally somewhat involved: Draw two Gaussian random maps,
$\eta_0$ and $\eta_1$, with zero mean and unit variance, and solve the
following equation for $\bar{\mathbf{s}}$,
\begin{equation}
(\mathbf{S}^{-1} + \mathbf{N}^{-1})\bar{\mathbf{s}} =
  \mathbf{N}^{-1}\mathbf{d} + \mathbf{L}^{-T} \eta_0 + \mathbf{N}^{-\frac{1}{2}},
\end{equation}
where $\mathbf{L}$ is the Cholesky decomposition of $\mathbf{S} =
\mathbf{L}\mathbf{L}^T$.  By multiplying both sides of this equation
with $(\mathbf{S}^{-1} + \mathbf{N}^{-1})^{-1}$, one immediately sees that
$\left<\bar{\mathbf{s}}\right> = \hat{\mathbf{s}}$, and a few more
computations show that
$\left<(\bar{\mathbf{s}}-\tilde{\mathbf{s}})(\bar{\mathbf{s}}-\tilde{\mathbf{s}})^T\right> = (\mathbf{S}^{-1} +
\mathbf{N}^{-1})^{-1}$, as required. 

For improved numerical stability, this linear system is in practice
rewritten into the following form,
\begin{equation}
(\mathbf{1} +
  \mathbf{L}^{T}\mathbf{N}^{-1}\mathbf{L})(\mathbf{L}^{-1}
  \bar{\mathbf{s}}) = \mathbf{L}^{T}\mathbf{N}^{-1}\mathbf{d} + \eta_0 + \mathbf{L}^{T}\mathbf{N}^{-\frac{1}{2}},
\label{eqsampling1}
\end{equation}
which is first solved for $\mathbf{x} = \mathbf{L}^{-1}
\bar{\mathbf{s}}$ by conjugate gradients, and then for
$\bar{\mathbf{s}} = \mathbf{L} \mathbf{x}$. For further
implementational details, see, e.g., \citet{eriksen:2008a}. Note,
however, that in previous papers equation (\ref{eqsampling1}) was always
written with symmetric signal covariance square roots,
$\mathbf{S}^{\frac{1}{2}} = (\mathbf{S}^{\frac{1}{2}})^T$. The current
form is based on the Cholesky decomposition, which is computationally
considerably cheaper than the symmetric form, especially for sparse
matrices.

Finally, we need a sampling algorithm for $P(\theta| \mathbf{s},
\mathbf{d})$. In previous publications, the main emphasis has been on
covariance matrices parametrized by the angular CMB power spectrum,
$C_{\ell m, \ell' m'} = C_{\ell} \delta_{\ell \ell'} \delta_{m m'}$. In
this case, $P(C_{\ell}|\mathbf{s},\mathbf{d})$ reduces to a simple
inverse Gamma distribution, for which there is a simple textbook
sampling algorithm available. As the details of this specific
algorithm is of little use for the application presented here, we
refer the interested reader to earlier papers for full details on this
procedure, e.g., \citet{wandelt:2004} or \citet{eriksen:2008c}.

\subsection{Gibbs sampling with non-diagonal covariances}

We now describe the two modifications to the CMB Gibbs sampler that
allow us to analyze models with non-diagonal covariances. This
involves adding support for non-diagonal covariance matrices for
$P(\mathbf{s} | \theta, \mathbf{d})$ and implementing a more general
sampling algorithm for $P(\theta| \mathbf{s}, \mathbf{d})$.

\subsubsection{Sampling from $P(\mathbf{s} | \theta,\mathbf{d})$}

We first consider sampling of sky maps, $\mathbf{s}$, given a set of
cosmological parameters, $\theta$, and the associated covariance
matrix $\mathbf{S}(\theta)$. Formally, the sampling algorithm for
$P(\mathbf{s} | \theta,\mathbf{d})$ is identical to that given by
equation (\ref{eqsampling1}). However, in this case $\mathbf{S}$ is a
non-diagonal matrix and the computational complexity is therefore
greatly increased. Only special cases can be considered, for instance
models that predict a sparse covariance matrix. This is the case for
the ACW model.

For general dense anisotropic covariance matrices, the memory
requirements scale as $\mathcal{O}(\ell_{\textrm{max}}^4)$,
effectively rendering studies of anisotropic models where
$\ell_{\textrm{max}} \gtrsim 100$ impossible.  However, working only with
sparse matrices, the memory consumption scales as
$\mathcal{O}(\ell_{\textrm{max}}^2)$, enabling calculations of
covariance matrices with $\ell_{\textrm{max}}$ well into the Planck
regime ($\ell_{\textrm{max}} \sim 2500$).

To be able to handle sparse matrices efficiently, we have ported the
LDL library of \citet{tdavis:2005} to Fortran 90, and incorporated
this into Commander. This library stores sparse matrices in a packed
format, and supports fast Cholesky decomposition. Our F90 version of
LDL may be obtained by sending an email to the authors, and will be
released publicly at a later time.

In the present paper, we are primarily concerned with the ACW model,
and the corresponding covariance matrix exhibits correlations between
$\ell$ and $\ell'=\{\ell,\ell\pm2\}$ and between $m$ and $m'=\{m,
m\pm1, m\pm2\}$. Thus, the number of elements up to
$\ell_{\textrm{max}}$ is $\mathcal{O}(15\ell_{\textrm{max}}^2)$.  For
example, for $\ell_{\textrm{max}} = 300$ the memory requirements are
$\sim 14$Mb with double precision complex numbers. Since the
covariance matrix is very sparse, the CPU time required for Cholesky
decomposition is nearly linear in $\ell_{\textrm{max}}^2$.

We define three different harmonic space limits in our code, namely
$\ell_{\textrm{max}}$, $\ell_{\textrm{low}}$ and
$\ell_{\textrm{high}}$. The former denotes the maximum multipole
moment of the full spherical harmonics composition used in the
analysis, while the latter two denotes the range in which the anisotropic
covariance matrix is used. In addition, we remove the monopole and
dipole from the analysis. Thus, the total covariance matrix reads
\begin{equation}
S_{\ell m, \ell' m'} =\left\{
\begin{array}{ll}
0 & \ell, \ell' \le 1 \\
C_{\ell}\delta_{\ell \ell'} \delta_{m m'}
& 2 \le \ell,\ell' < \ell_{\textrm{low}} \\
\frac{(C_{\ell}\delta_{\ell \ell'} \delta_{m m'} + \Delta_{\ell m, \ell'
  m'})}{1+g_*/3}
& \ell_{\textrm{low}} \le \ell, \ell' \le \ell_{\textrm{high}} \\
C_{\ell}\delta_{\ell \ell'} \delta_{m m'}
& \ell_{\textrm{high}} < \ell,\ell' \le \ell_{\textrm{max}}
\end{array}
\right.
\label{eq:full_covar}
\end{equation}
The reason for defining $\ell_{\textrm{min}}$ and
$\ell_{\textrm{max}}$ as free parameters is that it may be useful to
study the dependence of $\theta$ on a particular $\ell$-range.  On the
other hand, this implies that the model implemented in this paper is
only an approximation to the full ACW model, for which the
correlations extend over all $\ell$'s, and is only exact when
$\ell_{\textrm{high}} = \ell_{\textrm{max}}$.

\begin{figure}
\includegraphics[width=80mm]{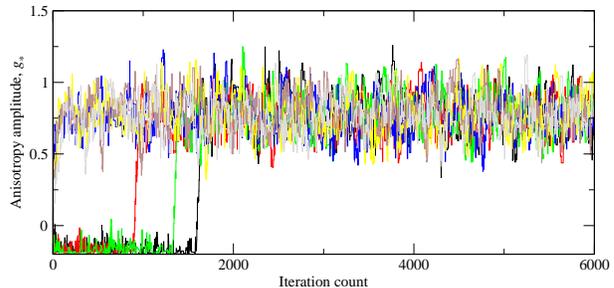}
\caption{Evolution of Gibbs chains mapping the posterior of a simulated
  data set. Note how chains trapped in the local maximum at
  negative anisotropy amplitude eventually converges to the positive maximum.}
\label{fig:burnin}
\end{figure}

With the sparse matrix operations implemented, the algorithm is
precisely the same as for the diagonal case, and both rely on the
solution of a linear system by Conjugate Gradients
\citep[CG;][]{eriksen:2004b}. In order to achieve an acceptable CG
convergence rate, it is therefore necessary to establish a good
preconditioner. However, as long as the off-diagonal elements remain
small, the standard diagonal covariance matrix preconditioner performs
reasonably even for the off-diagonal case. For the present paper, we
therefore adopt the same preconditioner as described by
\citet{eriksen:2004b}, which consists of the directly inverted full
matrix evaluated up to some $\ell_{\textrm{precond}}$, and then a
strictly diagonal matrix from $\ell_{\textrm{precond}}+1$ to
$\ell_{\textrm{max}}$.

The number of CG iterations per map making step is typically 70 for a
WMAP-type run, and with a total CPU time per iteration of about 15
seconds, the total cost for a single sample is $\sim20$ CPU
minutes. The average CPU time required to set up and perform a
Cholesky decomposition of the corresponding covariance matrix for
$\ell_{\textrm{max}} = 512$, $\ell_{\textrm{low}} = 2$ and
$\ell_{\textrm{high}} = 300$ is $\sim20$ seconds.

\subsubsection{Sampling from $P(\theta|\mathbf{s}, \mathbf{d})$}

Finally, we have to formulate a sampling algorithm for
$P(\theta|\mathbf{s}, \mathbf{d})$. Recall that for the diagonal power
spectrum case, this step is typically performed by a standard inverse
Gamma distribution sampler
\citep[e.g.,][]{gupta:2000,eriksen:2008c}. For the general case
considered here, we adopt a standard Metropolis MCMC sampler
\citep[e.g.,][]{liu:2001}.

\begin{figure*}
\includegraphics[width=170mm]{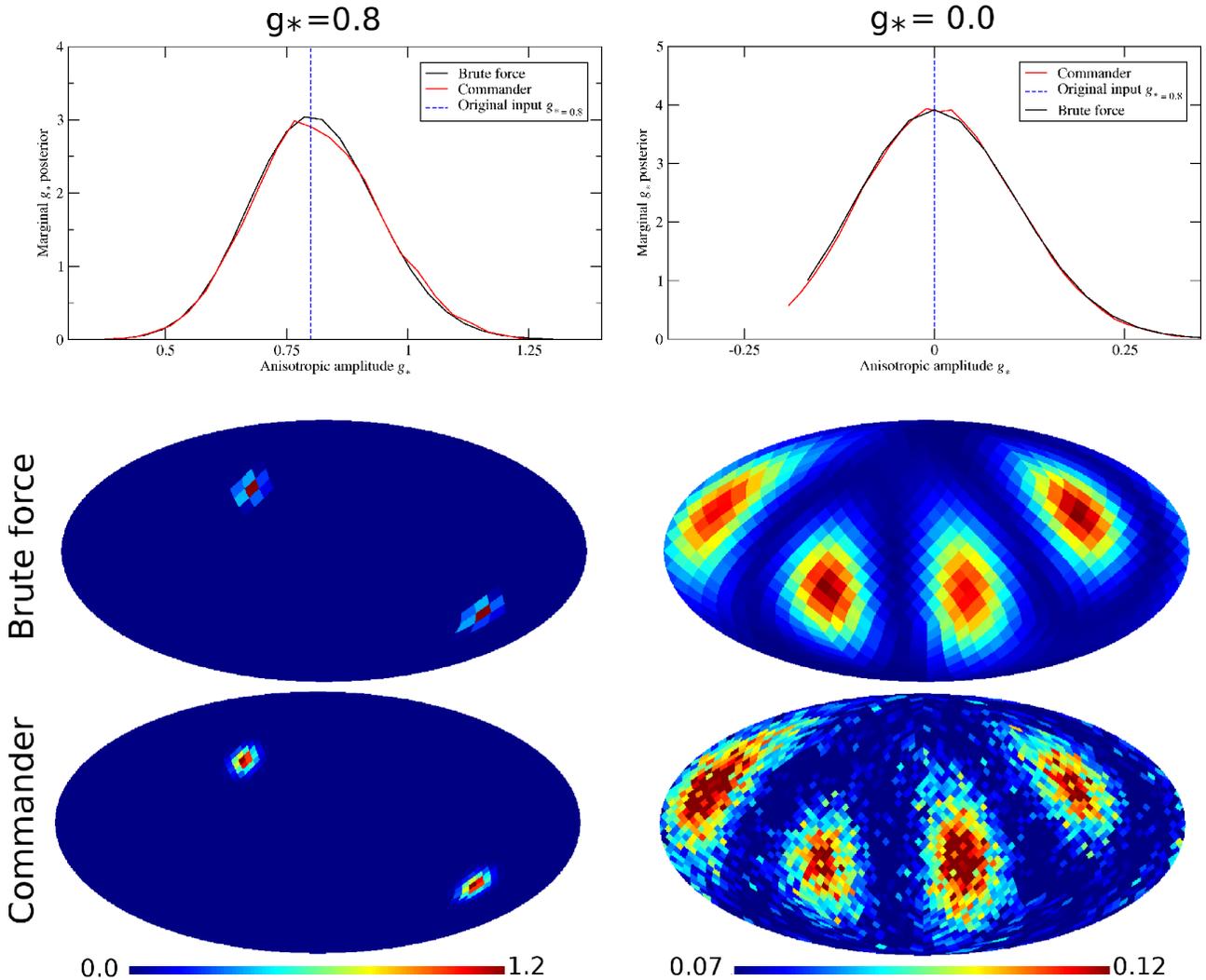}
\caption{Posterior distributions for simulated maps with a significant
  anisotropic amplitude $g_* = 0.8$ (left) and no anisotropic
  amplitude $g_* = 0.0$ (right). Note how the
  anisotropic input parameters $\theta, \phi, g_*$ were successfully reproduced.}
\label{fig:verification}
\end{figure*}

First, note that $P(\theta|\mathbf{s}, \mathbf{d}) =
P(\theta|\mathbf{s})$; if we already know the CMB sky perfectly, no
additional data can possibly tell us anything more about the
anisotropy parameters $\theta$. Second, although the CMB sky is now
manifestly anisotropic, we still assume that it is Gaussian, and the
target distribution therefore reads
\begin{equation}
  P(\theta|\mathbf{s}) \propto \frac{e^{-\frac{1}{2}\mathbf{s}^T \mathbf{S}^{-1}\mathbf{s}}}{\sqrt{|\mathbf{S}|}}.
\end{equation}
For sparse matrices, this may be directly evaluated by first computing
the Cholesky decomposition of $\mathbf{S} = \mathbf{LL}^t$, and then,
on the one hand, solve for $\mathbf{x} = \mathbf{L}\mathbf{s}$, and on
the other hand, compute $|\mathbf{S}| = |\mathbf{L}|^2$.

We adopt a simple symmetric proposal rule for the Metropolis sampler,
and the acceptance probability therefore simply reads
\begin{equation}
p = \frac{P(\theta^p|\mathbf{s})}{P(\theta^i|\mathbf{s})},
\end{equation}
where $\theta^p$ is the proposed sample and $\theta^i$ is the current
sample of the MCMC chain. Specifically, we adopt a Gaussian proposal
density for $g_*$ and a uniform proposal over a disk for
$\hat{\mathbf{n}}$, centered on the current state. The proposal
density is typically tuned by producing a short test chain before the
main run, such that the final observed acceptance rate lies between
0.2 and 0.7.

\begin{figure}
\includegraphics[width=80mm]{f7a.eps}
\includegraphics[width=80mm]{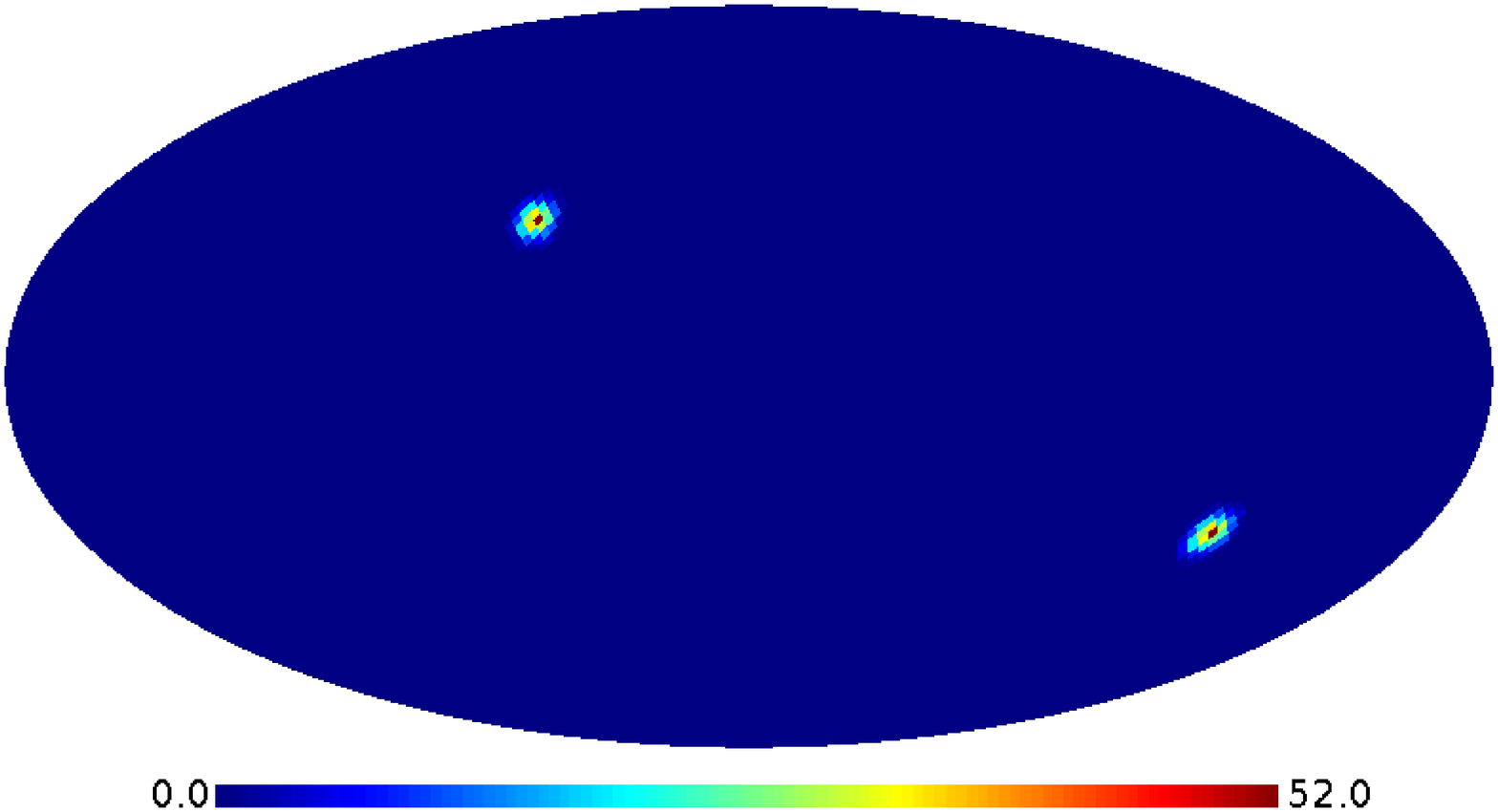}
\caption{
Posterior distributions for a simulated WMAP data set, using
  the V-band beam, V band RMS noise and the KQ85 sky cut. }
\label{fig:wmap_sim}
\end{figure}

Finally, because the computational cost is much lower for this step
than for $P(\mathbf{s}|\theta, \mathbf{d})$, we produce several
$\theta$ samples per main Gibbs iteration, to improve the convergence
properties of the chain. This essentially corresponds to performing a
partial Rao-Blackwellization \citep{chu:2005}. A typical number of
MCMC samples per main Gibbs iteration is 30.

\section{Applications to simulated data}
\label{sec:simulations}

We now apply the methods described above to simulated data, both in
order to validate the code and to build up intuition about the target
distribution. Note that the discussion from now on specializes
exclusively to the ACW model, and it is possible that other technical
issues than those described here may arise when considering other
models. Burn-in, mixing and convergence are issues that must be
considered on a case-to-case basis.

\subsection{Simulations}
\label{sec:sim_procedure}

To test our implementation and study the behavior of the algorithm in
general, we simulate a few different maps from the ACW model, and
analyze these maps with our modified Gibbs sampler. The CMB component
of these maps is made by generating a random vector, $\eta$, of
Gaussian uniform variates with zero mean and unit variance, and then
computing $\mathbf{s} = \mathbf{L}\eta$. This realization is then
convolved with a beam function and the HEALPix pixel window, before it
is projected on a HEALPix grid. Finally, Gaussian noise is added to
each pixel.

The first two simulations have a resolution of $N_{\textrm{side}} =
128$, $\ell_{\textrm{max}} = 256$, $\ell_{\textrm{low}} = 2$,
$\ell_{\textrm{high}} = 200$ and a Gaussian beam of $90'$ FWHM. The
noise RMS is $10\,\mu\textrm{K}$ uniformly over the full sky. The
first of the two simulations has an anisotropy amplitude of $g_* =
0.8$ and a preferred direction towards $(l,b) =
(57^{\circ},33^{\circ})$, and the other $g_* = 0$. These two
simulations are primarily used to compare the Gibbs sampler with
brute-force likelihood evaluation, which is only possible for uniform
noise and full-sky coverage.

Second, we generate a full WMAP5 like simulation based on the V1
differencing assembly (DA), with $g_* = 0.8$, $N_{\textrm{side}} =
512$, $\ell_{\textrm{max}} = 600$, $\ell_{\textrm{low}} = 2$,
$\ell_{\textrm{high}} = 300$ and beam and noise properties appropriate
for the V1 DA\footnote{http://lambda.gsfc.nasa.gov}. In this case, we
also apply the KQ85 sky cut \citep{gold:2008}, which removes 18\% of
the sky. This simulation is used to verify that correct results are
obtained for realistic WMAP data, including anisotropic (but
uncorrelated) noise and a sky cut.

\subsection{Burn-in and convergence}

We first consider the issue of burn-in and convergence, and analyze
the simulation with $g_* = 0.8$, uniform noise and full-sky
coverage. In Figure \ref{fig:burnin} we show the first 6000 $g_*$
samples produced by each of 14 chains. First, notice that the chains
immediately divide into two classes, one which converges quickly
towards $g_* \sim 0.8$ and one which hovers near the lower prior of
$g_* = -0.2$. This is due to the fact that the chains are initialized
randomly on the sphere, and those that happen to start close to the
non-physical local maximum (see Section \ref{sec:priors}) get
temporarily trapped in this local maximum. However, as the chains
explores the likelihood surface, they are able to converge into the
right regime, and find the correct value. In this case, all chains
have reached the equilibrium state after 1800 iterations. The
pre-burn-in samples must be rejected from the further analysis. For
now, we inspect each chain individually, to make sure that they have
all reached the common state. 

Note that there is a fundamental difference between low and high
signal-to-noise cases in this respect: If $g_*$ is low, the chains may
jump between local maxima, while if $g_*$ is high, some chains
typically start out in the global maximum and stay there, while others
start in the local maximum, and eventually converge into the right
regime. Which situation is relevant for a particular data set must be
considered on a case-by-case basis, by checking whether the chains
jump between states, or if they stay in one place. It is also
advisable to run many chains in parallel, randomly initialized over
the full sphere, to understand how many local maxima the distribution
has.

Second, once the chains have burned in, we must also ensure that they
collectively have converged to the full posterior. One possible
measure for this is the Gelman-Rubin $R$ statistic
\citep[e.g.,][]{gelman:1992, eriksen:2006}, which compares the
variances within a single chain with the variance between chains. If
the chains have converged properly, $R$ should be close to
unity. Typically, one recommends that $R$ should be less than 1.1 or
1.2. For the chains shown in Figure \ref{fig:burnin}, we find that $R
= 1.01$ after rejecting the first 2000 burn-in samples, indicating
very good convergence. Considering further subsets of these samples,
we find that 5000 samples is sufficient to achieve $R < 1.1$ and
20\,000 samples for $R < 1.02$. In the analyses presented later, we
always have more than 20\,000 samples.

\subsection{Validation}

We now analyze the two simulations with uniform noise and full-sky
coverage, having $g_* = 0$ and $g_* = 0.8$, respectively. In addition
to running the Gibbs sampler on these simulations, we also compute the
full three-dimensional likelihood function over a grid in $(g_*,
\hat{\mathbf{n}})$, and numerically integrate to produce brute-force
marginal posteriors.

The resulting distributions are shown in Figure
\ref{fig:verification}; the left column shows the $g_* = 0.8$ case,
and the right column shows the $g_* = 0$ case. We see, as expected,
that the two methods produce identical results, up to sampling
uncertainty and grid resolution. Note that this holds both for high
and low anisotropy amplitudes, indicating that the method is robust in
all regimes.

Next, we see that when the amplitude is large, there is only one
visible preferred direction in the direction posterior; the secondary
direction is too shallow to be seen. On the other hand, there are two
``preferred''directions in the $g_* = 0$ case. However, the span in
likelihood over the full sphere is in this case only a factor of two
between the least and most preferred directions, which essentially
indicates a uniform distribution.

In Figure \ref{fig:wmap_sim} we show similar plots for the WMAP
simulation with uncorrelated noise, based on the V1 differencing
assembly and $g_* = 0.8$. In this case it is not possible to evaluate
the likelihood directly, since the noise is inhomogeneous and there is
a sky cut. Still, we see that correct results are obtained. This
concludes the verification of both the method and our implementation,
and we are now ready to analyze the five-year WMAP temperature sky
maps.

\subsection{Forecasts for cosmic variance limited data}

Before turning to the analysis of the actual WMAP data, we compute the
uncertainty in $g_*$ as a function of $\ell_{\textrm{high}}$ for
full-sky noiseless data. (The lower limit is always kept at
$\ell_{\textrm{low}} = 2$.) This topic was also considered by
\citet{pullen:2007}, who presented both a more general formalism and
forecasts for specific experiments. Note, however, that our
parametrization is slightly different from theirs, as we introduce a
rescaling of the covariance matrix to eliminate the power spectrum
degeneracy (Section \ref{sec:degeneracy}).

We carry out this analysis by simulating anisotropic ACW maps with
$g_* = 0$ and different $\ell_{\textrm{high}}$, and analyze these with
the brute-force evaluation approach described above. No noise or beam
effects are included. For each case, we marginalize over
$\hat{\mathbf{n}}$ to obtain $P(g_*|\mathbf{d})$, and compute the
standard deviation, $\sigma(g_*)$, from this distribution.

\begin{figure}
\includegraphics[width=80mm]{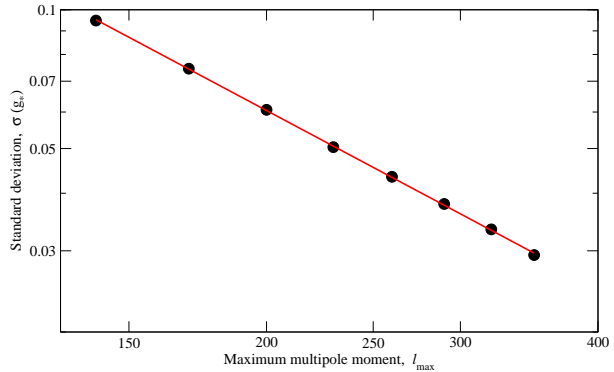}
\caption{Estimated uncertainty in $g_*$ as a function of
  $\ell_{\textrm{high}}$ (black dots) and a best-fit power law
  function (red line) for cosmic variance limited data.}
\label{fig:sigmas}
\end{figure}

Figure \ref{fig:sigmas} shows $\sigma(g_*)$ as a function of
$\ell_{\textrm{high}}$. From this figure, we see that $\sigma(g_*)$
is very close to a power law in $\ell_{\textrm{high}}$, in good
agreement with the arguments given by \citet{pullen:2007}. The
best-fit power law function is
\begin{equation}
\sigma(\ell_{\textrm{high}}; g_*) = 0.025\left(\frac{\ell_{\textrm{high}}}{400}\right)^{-1.27},
\end{equation}
and this can be used to produce rough forecasts for various
experiments. For instance, in this paper we conservatively adopt
$\ell_{\textrm{high}} = 400$ for the WMAP analysis, to avoid possibly
complicating high-$\ell$ issues such as point source confusion and
noise mis-estimation. In that case, we expect an uncertainty of
$\sigma(g_*) = 0.025$, before taking into account noise and sky
cut. This is in excellent agreement with the $\sigma(g_*) = 0.024$
result of \citet{pullen:2007}, derived with slightly different data
and model assumptions and a completely different approach.

\section{Application to the five-year WMAP data}
\label{sec:wmap}

We now analyze the five-year WMAP data, and present the full marginal
$P(g_*|\mathbf{d})$ and $P(\hat{\mathbf{n}}|\mathbf{d})$ posteriors for various
data cuts.

\subsection{Data}

In this paper, we consider the five-year WMAP temperature sky maps
\citep{hinshaw:2008}, and analyze the V- and W-bands (61 and 94 GHZ),
which are believed to be the cleanest WMAP bands in terms of residual
foregrounds. We adopt the template-corrected, foreground reduced maps
recommended by the WMAP team for cosmological analysis, and impose
both the KQ75 and KQ85 masks \citep{gold:2008}, which remove 28\% and
18\% of the sky, respectively. Point source cuts are imposed in both
cases.

We mainly analyze the data frequency-by-frequency, and consider the
combinations V1+V2 and W1 through W4. In addition, we compute the
posteriors for V1 and V2 separately. The noise RMS patterns and beam
profiles are taken into account for each DA individually. The noise is
assumed uncorrelated between pixels and bands. For details on joint
Gibbs analysis of multi-frequency data, see \citet{eriksen:2004b}. All
data used in this analysis are available from LAMBDA.

The angular resolutions of the V- and W-bands are $0.35^{\circ}$ and
$0.22^{\circ}$, respectively, and the sky maps are pixelized at a
HEALPix resolution of $N_{\textrm{side}} = 512$ with $7'$ pixels. We
therefore adopt a harmonic space cutoff of $\ell_{\textrm{max}} = 700$
and 800 for the two data sets, probing deeply into the noise dominated
regime. However, we never consider multipoles at $\ell > 400$ for the
anisotropic part of the signal covariance matrix, in order to
minimize the chance of systematic effects such as residual point
source contributions, beam uncertainties or noise mis-estimation to
affect our results. See Table \ref{tab:distributions} for a list of
the specific $\ell$-ranges considered.

We also note that the maps studied here are cleaned using external
templates \citep{gold:2008}, which must be considered a fairly rough
approach to foreground cleaning. A better approach is to use the joint
foreground and CMB Gibbs sampler \citep{eriksen:2008a}, which provides
the user with a CMB map marginalized over very general foreground
models. This work is currently underway for the five-year WMAP data,
and the results will be reported elsewhere (Dickinson et al., in
preparation). However, as an explicit foreground test we also analyze
the raw V-band data, from which no foreground templates have been
subtracted, and find very consistent results.

\subsection{Results}
\label{sec:results}

\begin{figure*}
\includegraphics[width=170mm]{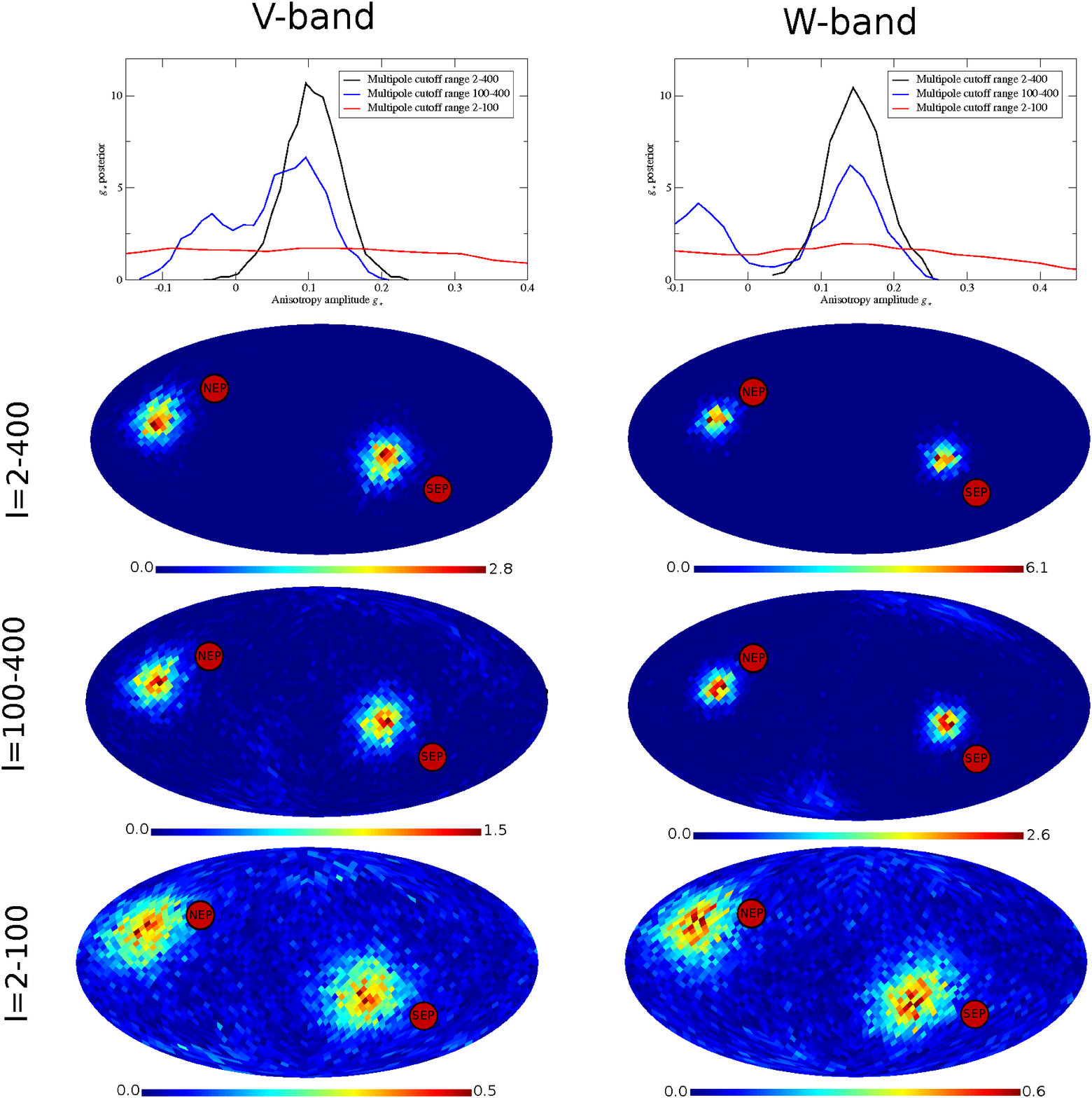}
\caption{Marginal ACW posteriors obtained from the V- (left) and
  W-band (right) WMAP temperature sky maps. Top row shows
  $P(g_*|\mathbf{d})$ and bottom three rows show
  $P(\hat{\mathbf{n}}|\mathbf{d})$ for three different
  $\ell$-ranges. Note the common preferred axis in both $\ell =
  [2,100]$ and $[100,400]$.}
\label{fig:wmap_posteriors}
\end{figure*}

We now present the marginal posteriors for the ACW model obtained from
the five-year WMAP temperature sky maps, as computed with the method
described in \S \ref{sec:method}. First, in the top row of Figure
\ref{fig:wmap_posteriors} we show the marginal anisotropy amplitude
posterior, $P(g_*|\mathbf{d})$, for V- (left column) and W-band (right
column), and in the three bottom rows we show the preferred direction
posteriors, $P(\hat{\mathbf{n}}|\mathbf{d})$. In Table
\ref{tab:distributions} the full set of results are summarized
quantitatively.

First, we see that there is an apparently clear detection of $g_* \ne
0$ when considering the full range of multipoles, $\ell =
[2,400]$. The W-band posterior has $g_* = 0.15 \pm 0.04$, nominally
corresponding to a $3.8\sigma$ detection, and the V-band posterior has
$g_* = 0.10 \pm 0.04$, internally consistent with W-band at
$\sim1 \sigma$. Second, the direction posteriors indicate a clearly
preferred direction pointing towards $(l,b) = (110^{\circ},
10^{\circ})$.

Further, this same direction is observed in both $\ell=[2-100]$ and
$\ell=[100,400]$, indicating that the structure is present over a
large range of angular scales. The results are also stable with
respect to sky cut, as the same pattern is seen with the KQ75 sky mask
as with the KQ85 cut, removing an additional 10\% of the sky.

\begin{deluxetable}{lcccc}
\tablewidth{0pt}
\tablecaption{Summary of marginal posteriors from WMAP5  \label{tab:distributions}} 
\tablecomments{In cases with no significant detection, the values
  for $g_*$ indicate the maximum posterior value and 95\% confidence
  regions. Otherwise, they indicate posterior mean and standard deviation.} 
\tablecolumns{4}
\tablehead{ Band & $\ell$ range  & Mask & Amplitude $g_*$ & Direction $(l,b)$ }
\startdata
V & $2-400$   &    KQ85    &   $0.10 \pm 0.04$               &  $(130^{\circ}, 10^{\circ})$                        \\ 
V & $100-400$ &    KQ85    &   $0.09  [ 0.084, 0.148]$           &  $(130^{\circ}, 10^{\circ})$                        \\ 
V & $2-100$   &    KQ85    &   $-0.07  [-0.156, 0.480]$           &  $(130^{\circ}, 15^{\circ})$                        \\ 
V & $2-400$   &    KQ75    &   $0.10  [-0.100, 0.158]$           &  $(130^{\circ}, 10^{\circ})$                        \\ 
V-raw & $2-400$ &  KQ85    &   $0.11 \pm 0.036$                  &$(130^{\circ}, 10^{\circ})$ \\
V1 & $2-400$  &    KQ85    &   $0.12 \pm 0.041$                   &$(130^{\circ}, 10^{\circ})$ \\
V2 & $2-400$  &    KQ85    &   $0.08 \pm 0.044$                   &$(130^{\circ}, 10^{\circ})$ \\
W & $2-400$   &    KQ85    &   $0.15 \pm 0.039$               &  $(110^{\circ}, 10^{\circ})$                        \\ 
W & $100-400$ &    KQ85    &   $0.14  [-0.097, 0.236]$           &  $(110^{\circ}, 10^{\circ})$                        \\ 
W & $2-100$   &    KQ85    &   $0.14  [-0.162, 0.470]$           & $(125^{\circ}, 20^{\circ})$                        
\enddata
\end{deluxetable}

\subsection{Sensitivity to systematics}

Given the nominally strong results found in the previous section, it
is imperative to search for possible systematic effects that might
explain the observations. In particularly, three major sources of
uncertainty should be considered in detail, namely non-cosmological
foregrounds, correlated noise and asymmetric beams. 

\begin{figure*}
\includegraphics[width=80mm]{f10a.eps}
\includegraphics[width=80mm]{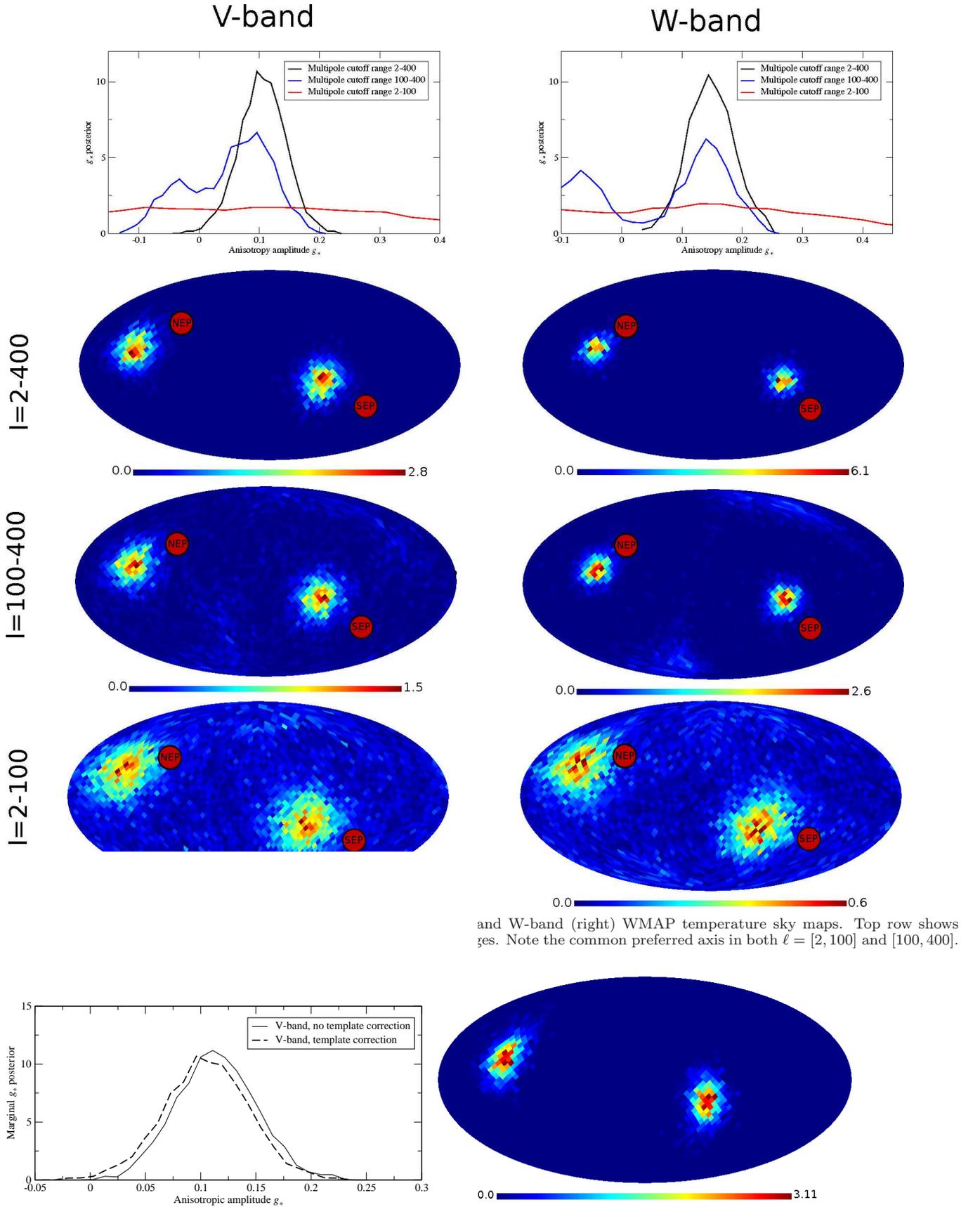}
\caption{Marginal ACW posteriors obtained from the non-template
  corrected V-band, $P(\hat n|\mathbf{d})$ (right) and $P(g_* |
  \mathbf{d})$ (left). Notice how $P(g_* | \mathbf d) $ is shifted
  insignificantly with respect to the template-corrected V-band posterior. }
\label{fig:imap_V}
\end{figure*}

First, residual Galactic foregrounds do not a priori appear as a
particularly promising candidate, given that the results are robust
with respect to both frequency and sky cut, and the preferred axis
does not point towards any natural Galactic direction.  Second, in
figure \ref{fig:imap_V} we show the posteriors obtained from the raw
V-band 5 year sky maps. It should be clear that galactic foregrounds
have little impact on these results.  Third, extragalactic point
sources do also not seem as a likely candidate, because the signature
is seen both at low and high $\ell$'s, and we never consider
multipoles above $\ell > 400$, precisely to avoid this type of
concerns.

The effect of correlated noise is harder to rule out. On the one hand,
returning to the simulated anisotropic CMB realization in Figure
\ref{fig:temperaturemaps}, we see that the main signature of the ACW
model is smoothed structures along the plane normal to the preferred
direction, and essentially no modifications along the preferred
direction. On the other hand, the main signature of correlated noise
is striping along the scan direction. 


Next, the ecliptic north pole has Galactic coordinates $(l,b) =
(96^{\circ}, 30^{\circ})$, which is $\sim 24^{\circ}$ ($32^{\circ}$)
away from the preferred direction for W-band (V-band) found in \S
\ref{sec:results}. The probability of obtaining such a close
 alignment by chance is $\sim10$\%, ($\sim16$\%) which is low enough
for correlated noise to be considered relevant for this particular
case.

To study the magnitude of this effect on $g_*$, we analyze realistic V
and W-band simulations with correlated noise. These noise simulations
were produced and published by the WMAP team in their 1-year data
release. To mimic realistic 5-year simulations, we coadd
five independent realizations for each differencing assembly. These
noise realizations are then added to an isotropic CMB sky realization,
and the sum is analyzed using the same procedure as in
\S \ref{sec:results}. We also analyze two single 1-year W4-band simulations,
which serve as a worst-case scenario, as the knee frequency of this
band is by far the highest of any WMAP DA, and the overall noise level
is higher by a factor of $\sim 4.5$ than the full 5-year W-band data.

\begin{figure*}
\includegraphics[width=170mm]{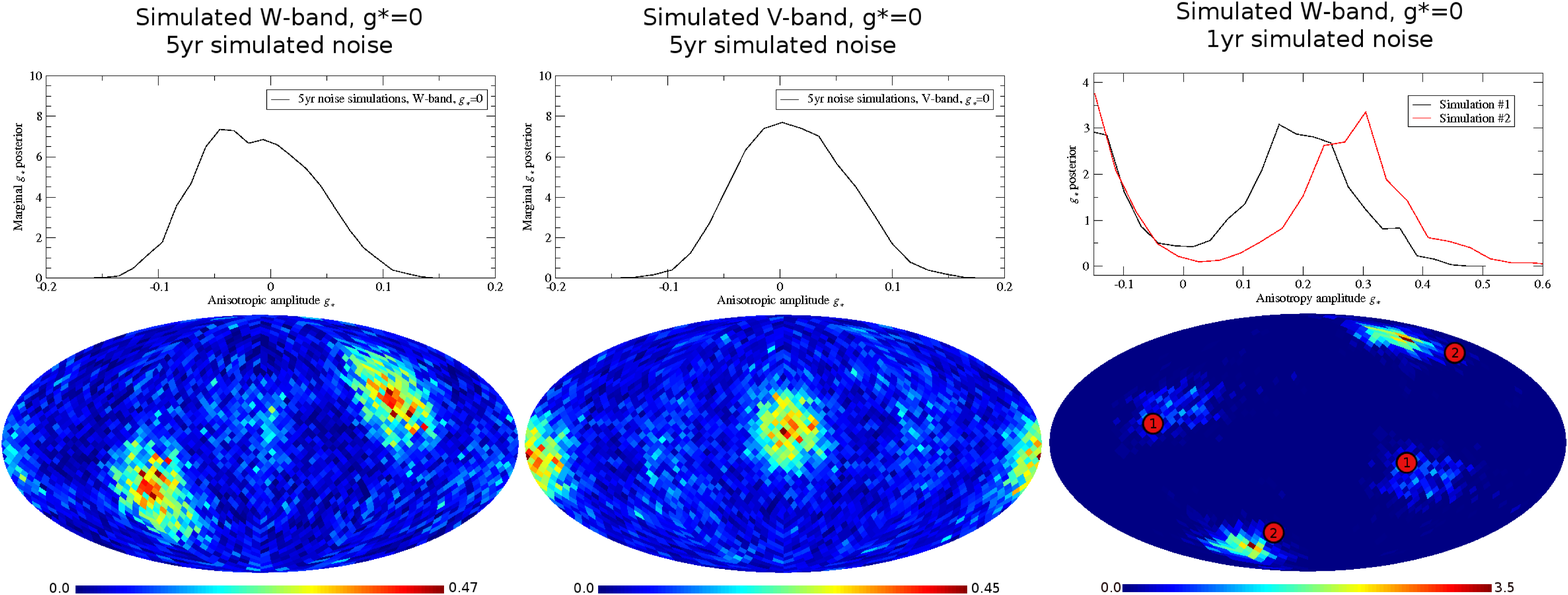}
\caption{Top: Posteriors for $g_*$ obtained from two 1-year WMAP W4
  simulations with correlated noise. Bottom: The preferred direction
  posterior for one of the two above simulations. The peak positions
  of the second simulations are indicated by red dots, marked by (1)
  and (2).}
\label{fig:correlated}
\end{figure*}

The results from these analyses are shown in Figure
\ref{fig:correlated}. The left and middle columns show the simulated
5-year posteriors for V and W-band, respectively, and the right column
shows the 1-year W4 posterior. The top row shows $P(g_*|\mathbf{d})$,
and the bottom panel shows $P(\hat{\mathbf{n}}|\mathbf{d})$.

First, note that with realistic 5-year noise no detection is made in
either the V- or W-band. Further, the peak sky position is different
in the V- and W-bands, and both have a very low significance. It
therefore seem unlikely that correlated noise can explain the results
found in \S \ref{sec:results}.

Still, caution is warranted, as the 1-year W4 posteriors do exhibit
traces of anisotropic contributions, with a peak amplitude larger than
the observed $g_*$ in the actual 5-year data. Yet, the match with the
structures observed in the real data is less than striking. First, the
correlated noise simulations show two independent peaks in the
directional posterior, while the WMAP data show one. Second, a
detailed study of the joint posteriors for the simulations show that
the peak along the ecliptic north pole corresponds to the
\emph{negative} peak in $P(g_*|\mathbf{d})$, while the WMAP data has a
positive $g_*$ along its preferred direction. Third, the preferred
axis found in the WMAP data are further away from the ecliptic pole
than the corresponding peak in the simulation posteriors.

Finally, in order to make a complete analysis, we should also consider
the impact of asymmetric beams. Ideally, one would prefer to address
this issue in the same manner as correlated noise, by analyzing
simulated CMB realizations with asymmetric beams. Unfortunately, we do
not have access to such simulations at this time, and it is difficult
to do a rigorous analysis. However, there are some arguments against
the asymmetric beams hypothesis. First, the effect is observed both at
low and high $\ell$'s, with very consistent positions. Second, the
observed preferred axis is $\sim$25--30 degrees away from the ecliptic
pole, and the posterior ratio of the ecliptic poles to the maximum
posterior is low. Finally, similar signatures are observed in both the
V and W bands, and in V1 and V2, which all have slightly different
beam patterns.

Nevertheless, at this point it would unwise to make strong claims
concerning a possible cosmological interpretation of the signature
found in \S \ref{sec:results}. Proper analysis of fully realistic
5-year WMAP simulations is required before one can attach cosmological
significance to these findings.

\section{Conclusions}
\label{sec:conclusions}

We have generalized a previously described CMB Gibbs sampler to allow
for exact Bayesian analysis of any anisotropic universe models that
predicts a sparse signal harmonic space covariance matrix. This
generalization involved incorporation of a sparse matrix library into
the existing Gibbs sampling code called ``Commander'', and
implementation of a new sampling algorithm for the anisotropy
parameters given a sky map, $P(\theta|\mathbf{s})$.

We then considered a special case of anisotropic universe models,
namely the \citet{ackerman:2007} model which generalizes the
primordial power spectrum $P(k)$ to include a dependence on direction,
$P(\mathbf{k})$. Explicit expressions for the resulting covariance
matrix is provided in their paper.

We implemented support for this model in our codes, and demonstrated
and validated the new tools with appropriate simulations. First, we
compared the results from the Gibbs sampler with brute-force
likelihood evaluations, and then verified that the input parameters
were faithfully reproduced in realistic WMAP simulations.

Finally, we analyzed the five-year WMAP temperature sky maps, and
presented for the first time the WMAP posteriors of the ACW model. The
results from this analysis are highly intriguing, but we emphasize
that the effect of instrumental systematics, particularly in the form
of correlated noise, must be better understood before the findings can
be given a cosmologically interpretation.

\begin{figure*}
\includegraphics[width=170mm]{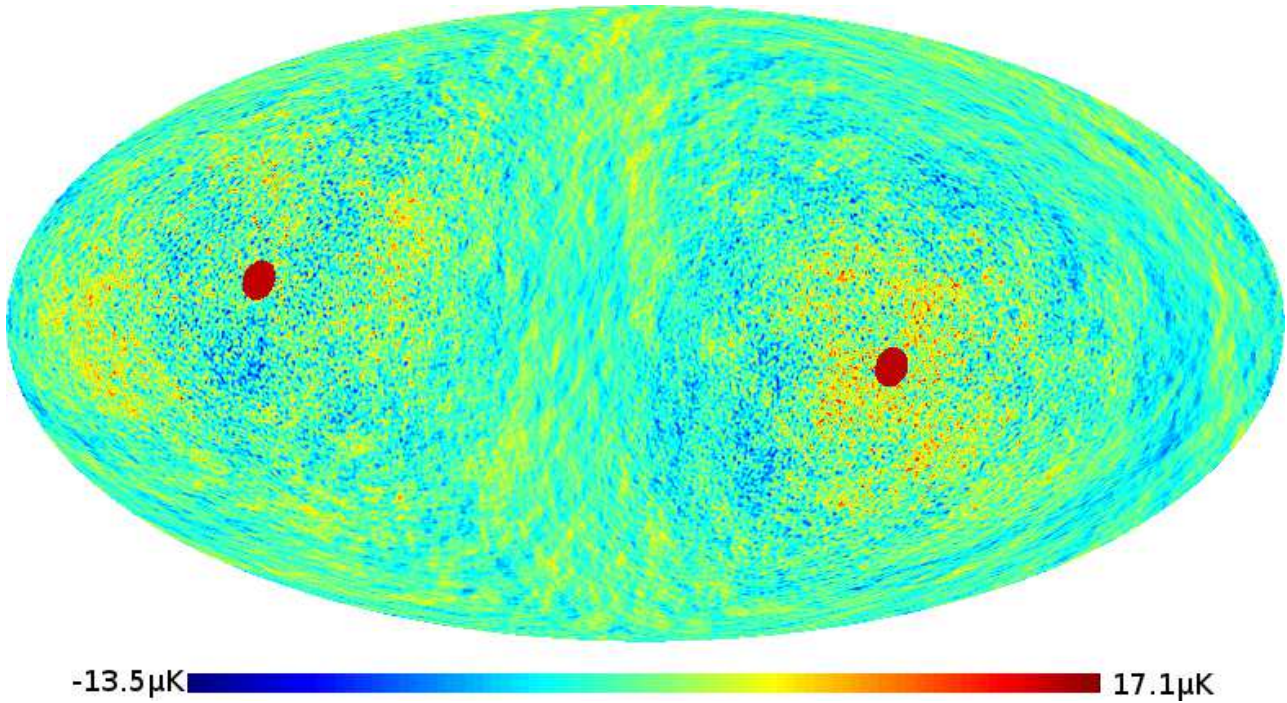}
\caption{A simulated realization drawn from a Gaussian distribution
  with zero mean and a covariance matrix given by the anisotropic
  $\Delta$ term in the ACW model, computed for an asymmetry amplitude
  of $g_* = 0.14$ and a preferred direction $(l,b) = (110^{\circ},
  10^{\circ})$, marked by red dots. Notice the rotational structure
  about the preferred direction. The amplitude of the anisotropic
  component is $\sim\pm 15\mu\textrm{K}$, or $\sim3$\% of the
  isotropic component.}
\label{figanisotropicmap}
\end{figure*}

Taken at face value, we find a preferred direction in the W-band WMAP
temperature data pointing towards $(l,b) = (110^{\circ}, 10^{\circ})$
(Galactic longitude and latitude), with an anisotropy amplitude of
$g_* = 0.15\pm 0.039$, formally corresponding to a $3.8\sigma$
detection of $g_* \ne 0$. Similar results for $g_*$ are found for the
V-band data, although with a somewhat lower significance ($g_* =
0.10 \pm 0.04$; $2.5\sigma$). The preferred direction is very stable
with respect to both data set and multipole range. Figure
\ref{figanisotropicmap} illustrates the underlying anisotropic
contribution for a simulation with parameters corresponding to the W
band posterior.

We have not been able to identify a plausible explanation for this
effect in terms of known systematics. First, foregrounds do not appear
to have much impact on the results, as consistent results are obtained
both from foreground-corrected and raw maps. Second, although
correlated noise does lead to a signature similar to the ACW model,
its amplitude appears too low in the 5-year data. The least well
constrained possibility is that of asymmetric beams, for which we lack
proper simulations. 

While this particular signature certainly is highly intriguing, we
would like to point out that the main purpose of this paper is the
demonstration of a general framework for analyzing anisotropic signal
models. This is useful both for studying particular universe models
(e.g., the ACW model), but also for understanding systematic effects
(e.g., correlated noise) in a given data set. We therefore believe
that these methods may be useful in a wide range of applications, only
some of which have been demonstrated in this paper.

\begin{acknowledgements}
  We thank Ned Wright for extremely useful feedback, and Tim Davis for
  helping out with the details with his sparse matrix LDL Cholesky
  decomposition library. We also thank Jeff Jewell, Frode K. Hansen,
  Magnus Axelsson and Kris G\'{o}rski for useful discussions. We
  acknowledge use of the HEALPix\footnote{http://healpix.jpl.nasa.gov}
  software \citep{gorski:2005} and analysis package for deriving the
  results in this paper. We acknowledge the use of the Legacy Archive
  for Microwave Background Data Analysis (LAMBDA). Support for LAMBDA
  is provided by the NASA Office of Space Science. The authors
  acknowledge financial support from the Research Council of Norway.
\end{acknowledgements}

\clearpage

\clearpage


\begin{thebibliography}{}





\bibitem[Ackerman et al.(2007)]{ackerman:2007} Ackerman, L., Carroll, 
S.~M., \& Wise, M.~B.\ 2007, \prd, 75, 083502 

\bibitem[Armendariz-Picon(2006)]{armendariz:2006} Armendariz-Picon, C.\ 
2006, Journal of Cosmology and Astro-Particle Physics, 3, 2 



\bibitem[Bennett et al.(2003)]{bennett:2003} Bennett, C. L., et al.\
  2003, \apjs, 148, 1

\bibitem[Chu et al.(2005)]{chu:2005} Chu, M., Eriksen, H.~K., Knox,
  L., G{\'o}rski, K.~M., Jewell, J.~B., Larson, D.~L., O'Dwyer, I.~J.,
  \& Wandelt, B.~D.\ 2005, \prd, 71, 103002

\bibitem[Cruz et al.(2005)]{cruz:2005} Cruz, M., 
Mart{\'{\i}}nez-Gonz{\'a}lez, E., Vielva, P., \& Cay{\'o}n, L.\ 2005, 
\mnras, 356, 29 

\bibitem[Davis(2005)]{tdavis:2005}
Davis, T.A, 2005,
ACM Trans. Math. Softw. vol 31 no 4


\bibitem[de Oliveira-Costa et al.(2004)]{de Oliveira-Costa:2004}
de Oliveira-Costa, A., Tegmark, M., Zaldarriaga, M., \& Hamilton,
A. 2004, \prd, 69, 063516

\bibitem[Erickcek et al.(2008)]{erickcek:2008} Erickcek, A.~L., 
Kamionkowski, M., 
\& Carroll, S.~M.\ 2008, PRL, submitted, [arXiv:0806.0377]

\bibitem[Eriksen et al.(2004a)]{eriksen:2004a} Eriksen, H.~K., Hansen, 
F.~K., Banday, A.~J., G{\'o}rski, K.~M., \& Lilje, P.~B.\ 2004a, \apj, 609, 1198 

\bibitem[Eriksen et al.(2004b)]{eriksen:2004b} Eriksen, H.~K., et al.\ 
2004b, \apjs, 155, 227 

\bibitem[Eriksen et al.(2006)]{eriksen:2006} Eriksen, H.~K., et al.\ 
2006, \apj, 641, 665 

\bibitem[Eriksen et al.(2007a)]{eriksen:2007a} Eriksen, H.~K., et al.\ 
2007a, \apj, 656, 641

\bibitem[Eriksen et al.(2007b)]{eriksen:2007b} Eriksen, H.~K., Huey, 
G., Banday, A.~J., G{\'o}rski, K.~M., Jewell, J.~B., O'Dwyer, I.~J., 
\& Wandelt, B.~D.\ 2007b, \apjl, 665, L1 

\bibitem[Eriksen et al.(2008a)]{eriksen:2008a} Eriksen, H.~K., Jewell, 
J.~B., Dickinson, C., Banday, A.~J., G{\'o}rski, K.~M., 
\& Lawrence, C.~R.\ 2008a, \apj, 676, 10 

\bibitem[Eriksen et al.(2008b)]{eriksen:2008b} Eriksen, H.~K., 
Dickinson, C., Jewell, J.~B., Banday, A.~J., G{\'o}rski, K.~M., 
\& Lawrence, C.~R.\ 2008b, \apjl, 672, L87 

\bibitem[Eriksen \& Wehus(2008)]{eriksen:2008c} Eriksen, H.~K., \&
  Wehus, I.\ K.\ 2008, ApJS, submitted, [arXiv:0806.3074]

\bibitem[Gelman \& Rubin(1992)]{gelman:1992}
Gelman, A., \& Rubin, D. 1992, Stat. Sci., 7, 457

\bibitem[Gold et al.(2008)]{gold:2008} Gold, B., et al.\ 2008, 
\apjs, in press, [arXiv:0803.0715]

\bibitem[G{\'o}rski et al.(2005)]{gorski:2005} 
  G{\' o}rski, K.~M., Hivon, E., Banday, A.~J., Wandelt, B.~D.,
  Hansen, F.\,K., Reinecke, M., \& Bartelmann, M. 2005, \apj, 622, 759


\bibitem[Emir G{\"u}mr{\"u}k{\c c}{\"u}oglu et 
al.(2007)]{contaldi:2007} Emir G{\"u}mr{\"u}k{\c c}{\"u}oglu, A.,
Contaldi, C.~R., 
\& Peloso, M.\ 2007, Journal of Cosmology and Astro-Particle Physics, 11, 5 


\bibitem[Gupta \& Nagar(2000)]{gupta:2000}
  Gupta, A.~K. \& Nagar, D.~K. 2000, Matrix Variate Distributions

\bibitem[Guth et al (1981)]{guth:1981}
Guth, A. H, 1981, \prd, 347

\bibitem[Hansen et al.(2004)]{hansen:2004} Hansen, F.~K., Banday, 
A.~J., \& G{\'o}rski, K.~M.\ 2004, \mnras, 354, 641 

\bibitem[Hinshaw et al.(2003)]{hinshaw:2003} Hinshaw, G., et al.\ 
2003, \apjs, 148, 63 

\bibitem[Hinshaw et al.(2007)]{hinshaw:2007} Hinshaw, G., et al.\ 
2007, \apjs, 170, 288 

\bibitem[Hinshaw et al.(2008)]{hinshaw:2008} Hinshaw, G., et al.\ 
2008, ApJ, submitted, [arXiv:0803.0732]

\bibitem[Jaffe et al.(2005)]{jaffe:2006} Jaffe, T.~R., Banday, 
A.~J., Eriksen, H.~K., G{\'o}rski, K.~M., \& Hansen, F.~K.\ 2005, \apjl, 
629, L1 

\bibitem[Jarosik et al.(2003)]{jarosik:2003} Jarosik, N., et al.\ 
2003, \apjs, 148, 29 

\bibitem[Jewell et al.(2004)]{jewell:2004}
Jewell, J., Levin, S., \& Anderson, C.~H.,
2004, \apj, 609 

\bibitem[Kanno et al.(2008)]{kanno:2008} Kanno, S., Kimura, M., 
Soda, J., \& Yokoyama, S.\ 2008, [arXiv:0806.2422] 



\bibitem[Komatsu et al.(2008)]{komatsu:2008} Komatsu, E., et al.\ 
2008, \apjs, in press, [arXiv:0803.0547]

\bibitem[Larson et al.(2007)]{larson:2007} Larson, D.~L., Eriksen, 
H.~K., Wandelt, B.~D., G{\'o}rski, K.~M., Huey, G., Jewell, J.~B., \& 
O'Dwyer, I.~J.\ 2007, \apj, 656, 653 

\bibitem[Lewis et al.(2000)]{lewis:2000} Lewis, A., Challinor, A., 
\& Lasenby, A.\ 2000, \apj, 538, 473 

\bibitem[Linde et al.(1982)]{linde:1982}
Linde, A. D., 1982, Phys. Lett. B 108, 389

\bibitem[Linde et al.(1983)]{linde:1983}
Linde, A. D., 1983, Phys. Lett. B 155, 295

\bibitem[Linde et al.(1994)]{linde:1994}
Linde, A. D., 1994, \prd 49, 748

\bibitem[Liu(2001)]{liu:2001} Liu, J. S., Monte Carlo Strategies in
  Scientific Computing, Cambridge, USA: Springer, 2001,   

\bibitem[Mukhanov et al.(1981)]{muhkanov:1981}
Muhkanov, V. F., Chibishov, G. V., \& Pis'mah Zh. 1981, Eskp. Teor. Fiz. 33, 549

\bibitem[O'Dwyer et al.(2004)]{odwyer:2004} O'Dwyer, I.~J., et al.\ 
2004, \apjl, 617, L99 

\bibitem[Oh et al.(1999)]{oh:1999} Oh, S.~P., Spergel, D.~N., 
\& Hinshaw, G.\ 1999, \apj, 510, 551 


\bibitem[Pullen \& Kamionkowski(2007)]{pullen:2007} Pullen, A.~R., \&
  Kamionkowski, M.\ 2007, \prd, 76, 103529

\bibitem[Ruhl et al (2003)]{ruhl:2003}
Ruhl et al., 2003, \apj 599, 786

\bibitem[Runyan et al.(2003)]{runyan:2003}
Runyan et al., 2003, \apj, J. Suppl. Ser. 149, 265

\bibitem[Scott et al.(2003)]{scott:2003}
Scott et al., 2003, \mnras 341, 1076

\bibitem[Smoot et al.(1992)]{smoot:1992}
Smoot et al., 1992, \apj 396, L1

\bibitem[Spergel et al.(2007)]{Spergel:2007} Spergel, D.~N., et al.\ 
2007, \apjs, 170, 377 


\bibitem[Starobinsky(1980)]{starobinsky:1980} Starobinsky, A.~A.,
1980, Physics Letters B, 91, 99 

\bibitem[Starobinsky et al (1982)]{starobinsky:1982}
Starobinsky, A. A., 1982, Phys. Lett. B 117, 175

\bibitem[Vielva et al.(2004)]{vielva:2004}
Vielva, P., Mart\'\i nez-Gonz\'alez, E., Barreiro, R. B., Sanz, J. L., \&
Cay\'on, L. 2004, \apj, 609, 22

\bibitem[Wandelt et al.(2004)]{wandelt:2004}
Wandelt, Benjamin D. and Larson, David L. and Lakshminarayanan, Arun 
\prd 70,8

\bibitem[Yokoyama \& Soda(2008)]{yokoyama:2008} Yokoyama, S., \&
  Soda, J.\ 2008, [arXiv:0805.4265]







\end{thebibliography}
\end{document}